\documentclass[11pt]{article}
\usepackage[margin=1in,left=1in]{geometry}
\usepackage{amsmath}
\usepackage{amsfonts}
\usepackage{amssymb}
\usepackage{amsthm}
\usepackage{mathtools}
\usepackage{graphicx}
\usepackage{color}
\usepackage{xcolor}
\usepackage{ucs}
\usepackage[utf8x]{inputenc}
\usepackage{xparse}
\usepackage{hyperref}
\usepackage[all]{xy}

\definecolor{aqua}{rgb}{0, 1.0, 1.0}
\definecolor{fuschia}{rgb}{1.0, 0, 1.0}
\definecolor{gray}{rgb}{0.502, 0.502, 0.502}
\definecolor{lime}{rgb}{0, 1.0, 0}
\definecolor{maroon}{rgb}{0.502, 0, 0}
\definecolor{navy}{rgb}{0, 0, 0.502}
\definecolor{olive}{rgb}{0.502, 0.502, 0}
\definecolor{purple}{rgb}{0.502, 0, 0.502}
\definecolor{silver}{rgb}{0.753, 0.753, 0.753}
\definecolor{teal}{rgb}{0, 0.502, 0.502}

\makeatletter
\newdimen\itex@wd%
\newdimen\itex@dp%
\newdimen\itex@thd%
\def\itexspace#1#2#3{\itex@wd=#3em%
\itex@wd=0.1\itex@wd%
\itex@dp=#2ex%
\itex@dp=0.1\itex@dp%
\itex@thd=#1ex%
\itex@thd=0.1\itex@thd%
\advance\itex@thd\the\itex@dp%
\makebox[\the\itex@wd]{\rule[-\the\itex@dp]{0cm}{\the\itex@thd}}}
\makeatother

\makeatletter
\newif\if@sup
\newtoks\@sups
\def\append@sup#1{\edef\act{\noexpand\@sups={\the\@sups #1}}\act}%
\def\reset@sup{\@supfalse\@sups={}}%
\def\mk@scripts#1#2{\if #2/ \if@sup ^{\the\@sups}\fi \else%
  \ifx #1_ \if@sup ^{\the\@sups}\reset@sup \fi {}_{#2}%
  \else \append@sup#2 \@suptrue \fi%
  \expandafter\mk@scripts\fi}
\def\tensor#1#2{\reset@sup#1\mk@scripts#2_/}
\def\multiscripts#1#2#3{\reset@sup{}\mk@scripts#1_/#2%
  \reset@sup\mk@scripts#3_/}
\makeatother

\makeatletter
\newbox\slashbox \setbox\slashbox=\hbox{$/$}
\def\itex@pslash#1{\setbox\@tempboxa=\hbox{$#1$}
  \@tempdima=0.5\wd\slashbox \advance\@tempdima 0.5\wd\@tempboxa
  \copy\slashbox \kern-\@tempdima \box\@tempboxa}
\def\slash{\protect\itex@pslash}
\makeatother

\def\clap#1{\hbox to 0pt{\hss#1\hss}}

\let\oldroot\root
\def\root#1#2{\oldroot #1 \of{#2}}
\renewcommand{\sqrt}[2][]{\oldroot #1 \of{#2}}

\DeclareSymbolFont{symbolsC}{U}{txsyc}{m}{n}
\SetSymbolFont{symbolsC}{bold}{U}{txsyc}{bx}{n}
\DeclareFontSubstitution{U}{txsyc}{m}{n}

\DeclareSymbolFont{stmry}{U}{stmry}{m}{n}
\SetSymbolFont{stmry}{bold}{U}{stmry}{b}{n}

\DeclareFontFamily{OMX}{MnSymbolE}{}
\DeclareSymbolFont{mnomx}{OMX}{MnSymbolE}{m}{n}
\SetSymbolFont{mnomx}{bold}{OMX}{MnSymbolE}{b}{n}
\DeclareFontShape{OMX}{MnSymbolE}{m}{n}{
    <-6>  MnSymbolE5
   <6-7>  MnSymbolE6
   <7-8>  MnSymbolE7
   <8-9>  MnSymbolE8
   <9-10> MnSymbolE9
  <10-12> MnSymbolE10
  <12->   MnSymbolE12}{}

\makeatletter
\def\re@DeclareMathSymbol#1#2#3#4{%
    \let#1=\undefined
    \DeclareMathSymbol{#1}{#2}{#3}{#4}}
\re@DeclareMathSymbol{\neArrow}{\mathrel}{symbolsC}{116}
\re@DeclareMathSymbol{\neArr}{\mathrel}{symbolsC}{116}
\re@DeclareMathSymbol{\seArrow}{\mathrel}{symbolsC}{117}
\re@DeclareMathSymbol{\seArr}{\mathrel}{symbolsC}{117}
\re@DeclareMathSymbol{\nwArrow}{\mathrel}{symbolsC}{118}
\re@DeclareMathSymbol{\nwArr}{\mathrel}{symbolsC}{118}
\re@DeclareMathSymbol{\swArrow}{\mathrel}{symbolsC}{119}
\re@DeclareMathSymbol{\swArr}{\mathrel}{symbolsC}{119}
\re@DeclareMathSymbol{\nequiv}{\mathrel}{symbolsC}{46}
\re@DeclareMathSymbol{\Perp}{\mathrel}{symbolsC}{121}
\re@DeclareMathSymbol{\Vbar}{\mathrel}{symbolsC}{121}
\re@DeclareMathSymbol{\sslash}{\mathrel}{stmry}{12}
\re@DeclareMathSymbol{\bigsqcap}{\mathop}{stmry}{"64}
\re@DeclareMathSymbol{\biginterleave}{\mathop}{stmry}{"6}
\re@DeclareMathSymbol{\invamp}{\mathrel}{symbolsC}{77}
\re@DeclareMathSymbol{\parr}{\mathrel}{symbolsC}{77}
\makeatother

\makeatletter
\def\Decl@Mn@Delim#1#2#3#4{%
  \if\relax\noexpand#1%
    \let#1\undefined
  \fi
  \DeclareMathDelimiter{#1}{#2}{#3}{#4}{#3}{#4}}
\def\Decl@Mn@Open#1#2#3{\Decl@Mn@Delim{#1}{\mathopen}{#2}{#3}}
\def\Decl@Mn@Close#1#2#3{\Decl@Mn@Delim{#1}{\mathclose}{#2}{#3}}
\Decl@Mn@Open{\llangle}{mnomx}{'164}
\Decl@Mn@Close{\rrangle}{mnomx}{'171}
\Decl@Mn@Open{\lmoustache}{mnomx}{'245}
\Decl@Mn@Close{\rmoustache}{mnomx}{'244}
\makeatother

\makeatletter
\DeclareRobustCommand\widecheck[1]{{\mathpalette\@widecheck{#1}}}
\def\@widecheck#1#2{%
    \setbox\z@\hbox{\m@th$#1#2$}%
    \setbox\tw@\hbox{\m@th$#1%
       \widehat{%
          \vrule\@width\z@\@height\ht\z@
          \vrule\@height\z@\@width\wd\z@}$}%
    \dp\tw@-\ht\z@
    \@tempdima\ht\z@ \advance\@tempdima2\ht\tw@ \divide\@tempdima\thr@@
    \setbox\tw@\hbox{%
       \raise\@tempdima\hbox{\scalebox{1}[-1]{\lower\@tempdima\box
\tw@}}}%
    {\ooalign{\box\tw@ \cr \box\z@}}}
\makeatother

\makeatletter
\NewDocumentCommand\mathraisebox{moom}{%
\IfNoValueTF{#2}{\def\@temp##1##2{\raisebox{#1}{$\m@th##1##2$}}}{%
\IfNoValueTF{#3}{\def\@temp##1##2{\raisebox{#1}[#2]{$\m@th##1##2$}}%
}{\def\@temp##1##2{\raisebox{#1}[#2][#3]{$\m@th##1##2$}}}}%
\mathpalette\@temp{#4}}
\makeatletter

\makeatletter
\def\udots{\mathinner{\mkern2mu\raise\p@\hbox{.}
\mkern2mu\raise4\p@\hbox{.}\mkern1mu
\raise7\p@\vbox{\kern7\p@\hbox{.}}\mkern1mu}}
\makeatother

\newcommand{\R}{\ensuremath{\mathbb R}}

\renewcommand{\(}{\begin{equation*}}
\renewcommand{\)}{\end{equation*}}
\newcommand{\bea}{\begin{eqnarray*}}
\newcommand{\eea}{\end{eqnarray*}}

\theoremstyle{italics}
\newtheorem{theorem}{Theorem}[section]

\theoremstyle{definition}

\newtheorem{example}[theorem]{Example}

\theoremstyle{remark}
\newtheorem{remark}[theorem]{Remark}
\newtheorem{note[theorem]}{Note}

\usepackage{amsfonts}

\newcommand{\Exterior}{\scalebox{.8}{\ensuremath \bigwedge}}

\begin{document}

\title{T-duality in rational homotopy theory via $L_\infty$-algebras} 

 \author{Domenico Fiorenza\thanks{Dipartimento di Matematica, La Sapienza Universit\`a di Roma, Piazzale Aldo Moro 2, 00185 Rome, Italy}\and 
     Hisham Sati\thanks{
Division of Science and Mathematics, New York University, Abu Dhabi, UAE}\and 
          Urs Schreiber\thanks{Mathematics Institute of the Academy, {\v Z}itna 25, 115 67 Praha 1, Czech Republic}
     }

\maketitle
\abstract{
We combine Sullivan models from rational homotopy theory with Stasheff's $L_\infty$-algebras to describe a duality in string theory. Namely, what in string theory is known as topological T-duality between $K^0$-cocycles in type IIA string theory and $K^1$-cocycles in type IIB string theory, or as Hori's formula, can be recognized as a Fourier-Mukai transform between twisted cohomologies when looked through the lenses of rational homotopy theory. We show this as an example of topological T-duality in rational homotopy theory, which in turn can be completely formulated in terms of morphisms of $L_\infty$-algebras.}

\tableofcontents

\section{Introduction} 

\medskip
A connected and simply connected space $X$ has a canonically defined based loop space $\mathbf{\Omega}X$, where the choice of the basepoint is irrelevant precisely due to the topological properties of $X$. From the space $\mathbf{\Omega}X$ one can reconstruct $X$ up to homotopy, as the classifying space for principal $\mathbf{\Omega}X$-fibrations, so the homotopy type of $X$ is completely known to the $\infty$-group $\mathbf{\Omega}X$. By analogy with the classical Lie group/Lie algebra correspondence, it should then be possible to reconstruct at least part of the homotopical content of $X$ from an infinitesimal version of  the $\infty$-group $\mathbf{\Omega}X$.  
One of the main result of rational homotopy theory \footnote{We are going to provide a very quick review of the basic ideas of rational homotopy theory in Section \ref{Sec-Basics}. See e.g. \cite{GM, FHT, FOT, FHT2, Hess06} for comprehensive surveys.} \cite{Quillen69} is that this rather vague statement can be rigorously formalized, and that a considerable amount of the homotopy type of $X$ is actually reconstructed: the rational homotopy type of $X$ is 
completely and faithfully encoded into a suitable $L_\infty$-algebra 
(\cite{SS} \cite{St} \cite{LS}) 
$\mathfrak{l}X$ 
which one may think of as being the 
infinitesimal version of the loop group $\mathbf{\Omega}X$; 
see \cite[Section 2]{BuijsFelixMurillo12} for a detailed account of this approach.

%
The \emph{semifree} DG-algebras of rational homotopy theory are then the Chevalley-Eilenberg 
algebras of these $L_\infty$-algebras. The $L_\infty$-algebra $\mathfrak{l}X$ can always be chosen to be concentrated in strictly negative degrees and with trivial differential, and these requirements determine $\mathfrak{l}X$ up to isomorphism. The corresponding Chevalley-Eilenberg 
algebras are the \emph{Sullivan model} DG-algebras of rational homotopy theory \cite{Sullivan77}. This can be summarized as follows:

\begin{center}
\begin{tabular}{|c|c|c|c|}
  \hline
  topological space & loop $\infty$-group &  $L_\infty$-algebra & Sullivan model
  \\
  \hline
  $X$ & $\mathbf{\Omega} X$ &  $\mathfrak{l}X$ & $\mathrm{CE}(\mathfrak{l}X)$
  \\
  \hline
\end{tabular}
\end{center}

The differential graded commutative algebra $A_X=\mathrm{CE}(\mathfrak{l}X)$ is then in turn directly related to the geometry of $X$ via the de Rham complex\footnote{We will be mostly concerned with smooth manifolds and so we will usually work over the field $\mathbb{R}$ of real numbers; one can more generally work over a characteristic zero field $\mathbb{K}$ by replacing de Rham complex of smooth differential forms with the de Rham complex of piecewise polynomial differential forms with coefficients in $\mathbb{K}$ associated with a simplicial set whose topological realization is homotopy equivalent to $X$, see \cite{Sullivan77}.} $\Omega^\bullet(X)$; namely, $A_X$ comes equipped with a quasi-isomorphism of  DGCAs $A_X\to \Omega^\bullet(X)$. 
This gives a direct connection to the notion of Lie algebroid cocycles on smooth manifolds, since, if $X$ is a smooth manifold and $\mathfrak{A}$ is a Lie algebroid, a $\mathfrak{A}$-valued cocycle on $X$ is by definition a morphism of Lie algebroids $TX\to \mathfrak{A}$, and so, equivalently, a morphism of DGCAs
\[
\mathrm{CE}(\mathfrak{A})\longrightarrow \Omega^\bullet(X).
\]
When $\mathfrak{A}=TY$, the tangent Lie algebroid of another manifold $Y$, by abuse of notation we call $Y$-valued cocycles the $TY$-valued cocycles on $X$, i.e., the DGCA morphisms $\Omega^\bullet(Y)\to \Omega^\bullet(X)$. In particular, every smooth morphism between $X$ and $Y$ naturally induces an $Y$-valued cocycle on $X$ and every
$Y$-valued cocycle on $X$ is of this form. Indeed, any morphism of DGCAs $\varphi\colon \Omega^\bullet(Y)\to \Omega^\bullet(X)$ induces in particular a morphism of commutative algebras $\Omega^0(Y)\to \Omega^0(X)$ and so at the level of degree zero components the morphism $\varphi$ is the pullback along a smooth map $f\colon X\to Y$.\footnote{This is sometimes known as  the Milnor's exercise; see 
\cite[Lemma 35.8; Corollaries 35.9, 35.10]{kms} for a proof.} Since $\Omega^\bullet(Y)$ is generated by $\Omega^0(Y)$ as a differetial graded commutative algebra, this implies that $\varphi=f^*$ in every degree. In other words we see that if $Y$ is a smooth manifold, then $Y$-valued cocycles on $X$ are precisely smooth maps $X\to Y$.

\medskip
This suggests the following immediate generalization: if $\mathfrak{l}Y$ is a 
Sullivan model for a smooth manifold $Y$, a smooth map $X\to \mathfrak{l}Y$ is by definition a DGCA morphism
\[
\mathrm{CE}(\mathfrak{l}Y)\longrightarrow \Omega^\bullet(X).
\]
By definition of Sullivan minimal model, $\mathrm{CE}(\mathfrak{l}Y)$ is a free polynomial algebra on certain generators $\{x_{\alpha_1},\dots x_{\alpha_k}\}$, with a differential which therefore will have the form
\[
dx_{\alpha_i}=P_{\alpha_i}(x_{\alpha_1},\dots x_{\alpha_k})\;.
\]
for some polynomial $P_{\alpha_i}$. Consequently, we see that a smooth map $X\to \mathfrak{l}Y$ is equivalently the datum of a collection of differential forms $\omega_{\alpha_i}$ on $X$ such that
\begin{equation}\label{eq.system}
d\omega_{\alpha_i}=P_{\alpha_i}(\omega_{\alpha_1},\dots\omega_{\alpha_k}),
\end{equation}
where now $d$ is the de Rham differential and the product is the wedge product of differential forms. Read the other way round, this says that every system of differential equations of the form (\ref{eq.system}) can be seen as a smooth map to a real Sullivan model. In particular, a field theory whose fields are differential forms obeying equations of the form (\ref{eq.system}) can be interpreted as a $\sigma$-model type field theory, with target space given by a real Sullivan model. All this immediately generalizes to the case of a smooth supermanifold $X$.

\medskip
An interesting example is provided by the fields usually denoted $G_4$ and $G_7$ in M-theory. Namely, such a pair of fields is naturally identified with the datum of a 4-form and a 7-form on 
spacetime $X$ with $dG_4=0$ and $dG_7=G_4\wedge G_4$, see \cite{Du}. As emphasized in \cite{top} \cite{tcu}, the Sullivan model of the 4-sphere over $\mathbb{R}$ is the polynomial algebra 
$\mathbb{R}[x_4,x_7]$ on two generators $x_4$ and $x_7$ in degree 4 and 7, respectively, and with differential given by $dx_4=0$ and $dx_7={x_4}^2$, so that the pair $(G_4,G_7)$ is precisely the datum of a smooth map from the smooth (super-)manifold $X$ 
to $\mathfrak{l}S^4$ and M-theory is consequently seen as a $\sigma$-model with values in $\mathfrak{l}S^4$.
Remarkably, in the Chevalley-Elenberg algebra of the superMinkowski space $\R^{10, 1|{\bf 32}}$, one has a degree $(4|0)$ element $g_4$ corresponding to what is called the $C$-field in M-theory and a degree 
$(7|0)$ element $g_7$ (called the dual of the $C$-field) which satisfy $dg_4=0$ and  $dg_7=g_4^2$, so that they define a map  $\R^{10, 1|{\bf 32}} \to  \mathfrak{l}S^4$. This implies that every worldvolume in the spacetime $\R^{10, 1|{\bf 32}}$ is naturally equipped with a map to 
$\mathfrak{l}S^4$, and so with M-theory fields, by restriction. 
%

\medskip
The superMinkowski space $\R^{10,1|{\bf 32}}$ behaves, from the point of view of rational homotopy theory, as a principal $U(1)$-bundle over the superMinkowski space $\mathbb{R}^{9,1\vert \mathbf{16}+ \overline{\mathbf{16}}}$. The M-theory morphism $\R^{10, 1|{\bf 32}}\to \mathfrak{l}S^4$ considered above then leads to considering the following geometric situation: a principal $U(1)$-bundle $P\to M$ together with a smooth map $P\to Y$, for some space $Y$.  
The total space $P$ is the homotopy fiber of the classifying map $M\to BU(1)$ for the bundle, 
and the general reduction in this case is described in \cite{MS}.
The homotopy fiber functor has a right adjoint, called ``cyclification'', mapping a space $Y$ to the \emph{twisted loop space} $\mathrm{cyc}(Y)=\mathcal{L}Y/\!/U(1)$, given by the homotopy quotient of the free loop space of $Y$ by the rotation of loops action: 
\[
\xymatrix{
\mathrm{spaces}\ar@/^1.6pc/[rr]^{\mathrm{cyc}}&&
\mathrm{spaces}/BU(1)\ar@/^1.6pc/[ll]^{\mathrm{hofib}}.
}
\]
The smooth map $P\to Y$ will, therefore, be equivalent to the datum of a smooth map $M\to \mathrm{cyc}(Y)$. This topological construction, capturing what in the physics literature is known as ``double dimensional reduction'', 
%
%
immediately translates to the rational homotopy theory/$L_\infty$-algebra setting, where we 
find an adjunction
\[
\xymatrix{
L_\infty\mathrm{\text{-}algebras}\ar@/^1.6pc/[rr]^{\mathrm{cyc}}&&L_\infty\mathrm{\text{-}algebras}/b\mathfrak{u}_1\ar@/^1.6pc/[ll]^{\mathrm{hofib}}.
}
\]
When applied to the $\mathfrak{l}S^4$-valued cocycle on $\R^{10,1|{\bf 32}}$, this produces a $\mathrm{cyc}(\mathfrak{l}S^4)$-valued cocycle on $\mathbb{R}^{9,1\vert \mathbf{16}+ \overline{\mathbf{16}}}$, which can be identified with (part of the data of) a twisted even K-theory cocycle. This corresponds to the double dimensional reduction from M-brane charges in 11d to string and  brane charges in 10d type IIA string theory. In particular, one recovers this way the string IIA twisted $K^0$-cocycles of \cite{CAIB00}. See \cite{FSS16a} for details.  
Note that while we have chosen to use rational homotopy theory to describe T-duality, 
at times using rational spectra is not a choice but rather a condition -- see \cite{LSW} 
for a precise and general statement. 

\medskip
The superMinkowski space $\mathbb{R}^{9,1\vert \mathbf{16}+ \overline{\mathbf{16}}}$ is in turn, again from the point of view of rational homotopy theory, a principal $U(1)$-bundle over the superMinkowski space $\mathbb{R}^{8,1|\mathbf{16}+\mathbf{16}}$ and, as such, it is classified by a 2-cocycle $c_2^{\mathrm{IIA}}$ in the (super-)Chevalley-Eilenberg algebra of $\mathbb{R}^{8,1|\mathbf{16}+\mathbf{16}}$. Quite remarkably, $\mathrm{CE}(\mathbb{R}^{8,1|\mathbf{16}+\mathbf{16}})$ carries also another, independent, 2-cocycle $c_2^{\mathrm{IIB}}$, corresponding to the superMinkowski space $\mathbb{R}^{9,1\vert \mathbf{16}+ \mathbf{16}}$. Moreover, the product $c_2^{\mathrm{IIA}}c_2^{\mathrm{IIB}}$ is an exact 4-cochain with an explicit trivializing 3-cochain. Thus, the pair of superMinkowski spaces $(\mathbb{R}^{9,1\vert \mathbf{16}+ \overline{\mathbf{16}}},\mathbb{R}^{9,1\vert \mathbf{16}+ \mathbf{16}})$ realizes in rational homotopy theory the data of a topological T-duality configuration \cite{BunkeSchick05}. As a consequence, one can bijectively transfer twisted $K^0$-cocycles in type IIA string theory to $K^1$-cocycles in type IIB string theory. In particular the string IIA twisted $K^0$-cocycles of \cite{CAIB00} are transformed into the string IIA twisted $K^1$-cocycles of \cite{IIBAlgebra}. This phenomenon, known as rational topological T-duality and explicitly expressed by the Hori's formula \cite{Ho}, can be formally derived by the properties of the $L_\infty$-algebra $b\mathfrak{tfold}$, providing the rational homotopy theory description of the universal space for T-duality, see \cite{FSS16}. 
Here we emphasize an aspect that remains somehow hidden in the exposition given in \cite{FSS16}. Namely, that the Hori's formula is precisely a Fourier-Mukai transform in the context of twisted $L_\infty$-algebra cohomology.
See \cite{Hu} for general background on Fourier-Mukai transforms and 
also \cite{BEM} \cite{AHR} \cite{Ru}   
for other discussions  in the context of T-duality.


\section{Dimensional reduction in rational homotopy theory}

%
\subsection{Twisted de Rham cohomology and twisted Fourier-Mukai transforms}
\label{Sec-twdR}

In order to prepare for the kind of construction we are going to describe in the setting of $L_\infty$-algebras, let us first recall its classical geometric counterpart: the Fourier-Mukai transform in twisted de Rham cohomology. 
Let therefore $X$ be a smooth manifold. One can twist the de Rham differential $d\colon \Omega^\bullet(X;\mathbb{R})\xrightarrow{d}\Omega^{\bullet}(X;\mathbb{R})$
by a 1-form $\alpha$, defining the twisted de Rham operator $d_\alpha\colon \Omega^\bullet(X;\mathbb{R})\xrightarrow{d}\Omega^{\bullet}(X;\mathbb{R})$ as $d_\alpha\omega=d\omega+\alpha\wedge\omega$. The operator $d_\alpha$ does not square to zero in general: $d_\alpha^2$ is the multiplication by the exact 2-form $d\alpha$. This means that precisely when $\alpha$ is a closed 1-form, the operator $d_\alpha$ is a differential, defining an $\alpha$-twisted de Rham complex $(\Omega^\bullet(X),d_\alpha)$. The cohomology of this complex is called the  $\alpha$-twisted de Rham cohomology of $X$ and it will be denoted by the symbol $H_{\mathrm{dR};\alpha}^\bullet(X)$. 

\medskip
The operator $d_\alpha$ is a connection on the trivial $\mathbb{R}$-bundle over $X$, which is flat precisely when $\alpha$ is closed. This means that for a closed 1-form $\alpha$, the $\alpha$-twisted de Rham cohomology of $X$ is actually a particular instance of flat cohomology or cohomology with local coefficients. This point of view is discussed extensively in \cite{GS1}. 
Having identified $d_\alpha$ with a connection, it is natural to think of gauge transformations as the natural transformations in twisted de Rham cohomology. More precisely, since we are in an abelian setting with a trivial $\mathbb{R}$-bundle, two connections $d_{\alpha_1}$ and $d_{\alpha_2}$ will be gauge equivalent exactly when there exists a smooth function $\beta$ on $X$ such that $\alpha_1=\alpha_2+d\beta$, i.e., when the two closed 1-forms $\alpha_1$ and $\alpha_2$ are in the same cohomology class. When this occurs, the two twisted de Rham complexes $(\Omega^\bullet(X),d_{\alpha_1})$ and $(\Omega^\bullet(X),d_{\alpha_2})$ are isomorphic, with an explicit isomorphism of complexes given by the multiplication by the smooth function $e^\beta$. That is, if $\omega$ is a differential form on $X$, we have
\begin{align*}
d_{\alpha_2}(e^\beta \wedge \omega)&=d(e^\beta\wedge \omega)+\alpha_2\wedge e^\beta\wedge \omega \\
&=d\beta\wedge e^\beta\wedge \omega+e^\beta\wedge d\omega+\alpha_2\wedge e^\beta\wedge\omega \\
&=e^\beta\wedge\left((d\beta+\alpha_2)\wedge \omega+ d\omega\right) \\
&=e^\beta\wedge\left(\alpha_1\wedge \omega+ d\omega\right) \\
&=e^\beta\wedge d_{\alpha_1}\omega\;.
\end{align*}
In particular, multiplication by $e^\beta$ induces an isomorphism in twisted cohomology
\[
e^\beta\colon H_{\mathrm{dR};\alpha_1}^\bullet(X)\overset{\sim}{\longrightarrow} H_{\mathrm{dR};\alpha_2}^\bullet(X)\;.
\]
We now investigate the functorial behavior of twisted cohomology with respect to a smooth map $\pi\colon Y\to X$. It is immediate to see that, since the pullback morphism $\pi^*\colon \Omega^\bullet(X)\to \Omega^\bullet(Y)$ is a morphism of DGCAs, it induces a morphism of complexes
\[
\pi^*\colon (\Omega^\bullet(X),d_\alpha)
\longrightarrow
 (\Omega^\bullet(Y),d_{\pi^*\alpha})\;.
\]
In turn this gives a pullback morphism in twisted cohomology
\[
\pi^*\colon H_{\mathrm{dR};\alpha}^\bullet(X)
\longrightarrow
 H_{\mathrm{dR};\pi^*\alpha}^\bullet(Y)\;.
\]
The pushforward morphism is a bit more delicate. To begin with, given a smooth map $\pi\colon Y\to X$ we in general have no pushforward morphism of complexes $\pi_*\colon \Omega^\bullet(Y)\to \Omega^\bullet(Y)$. However we do have such a morphism of complexes, up to a degree shift, if $Y\to X$ is not a general smooth map but it is an oriented fiber bundle with typical fiber $F$ which is a compact closed oriented manifold. In this case $\pi_*$ is given by integration along the fiber and is a morphism of complexes $\pi_*\colon 
\big(\Omega^\bullet(Y),d\big)\to \big(\Omega^\bullet(X)[-\dim F],d[-\dim F]\big)$. Yet, $\pi_*$ will not induce a morphism 
$\pi_*\colon \big(\Omega^\bullet(Y),d_\alpha\big)\to 
\big(\Omega^\bullet(X)[-\dim F],d_{\pi_*\alpha}[-\dim F]\big)$, and actually a minute's reflection reveals that the symbol $d_{\pi_*\alpha}$ just makes no sense. However, when $\alpha$ is not just a generic 1-form on $Y$ but it is a 1-form pulled back from $X$, then everything works fine. Namely, the projection formula
\[
\pi_*(\pi^*\alpha\wedge \omega) = (-1)^{\deg \alpha \dim F}\alpha\wedge \pi_*\omega
\]
precisely says that $\pi_*$ is a morphism of chain complexes
\[
\pi_*\colon (\Omega^\bullet(Y),d_{\pi^*\alpha})
\longrightarrow
\big(\Omega^\bullet(X)[-\dim F],d_{\alpha}[-\dim F]\big)
\]
and so it induces a pushforward morphism in twisted cohomology
\[
\pi^*\colon H_{\mathrm{dR};\pi^*\alpha}^\bullet(Y)
\longrightarrow
H_{\mathrm{dR};\alpha}^{\bullet-\dim F}(X).
\]

\paragraph{Fourier-Mukai transforms in twisted de Rham cohomology.}
All of the above suggests as to cook up a Fourier-type transform in twisted cohomology. Assume we are given a span of smooth manifolds
\[
\xymatrix@R=1.5em{
& Y\ar[dl]_{\pi_1}\ar[dr]^{\pi_2}&\\
X_1&&X_2,
}
\]
with $Y\xrightarrow{\pi_2} X_2$ an oriented fiber bundle with compact closed oriented fibers. Let $\alpha_i$ be a closed 1-form on $X_i$, and assume that the two 1-forms $\pi_1^*\alpha_1$ and  $\pi_2^*\alpha_2$ are cohomologous in $Y$, with $\pi_1^*\alpha_1-\pi_2^*\alpha_2= d\beta$. Then we have the sequence of morphisms of chain complexes
\[
(\Omega^\bullet(X_1),d_{\alpha_1})\xrightarrow{\pi_1^*} (\Omega^\bullet(Y),d_{\pi_1^*\alpha_1})\xrightarrow{e^\beta} (\Omega^\bullet(Y),d_{\pi_2^*\alpha_2})
\xrightarrow{\pi_{2*}} 
\big(\Omega^\bullet(X_2)[-\dim F_2],d_{\alpha_2}[-\dim F_2]\big)
\]
whose composition defines the Fourier-Mukai transform with kernel $\beta$ in twisted de Rham cohomology
\[
\Phi_\beta\colon H_{\mathrm{dR};\alpha_1}^\bullet(X_1) 
\longrightarrow
 H_{\mathrm{dR};\alpha_2}^{\bullet-\dim F_2}(X_2)\;.
\]
Writing ``$\int_F$" for $\pi_{2*}$ and writing ``$\cdot$" for the right action of $\Omega^\bullet(X)$ on $\Omega^\bullet(Y)$ given by $\eta\cdot\omega=\eta\wedge \pi_1^*\omega$ makes it evident why this is a kind of Fourier transform
\[
\Phi_\beta\colon \omega\longmapsto \int_{F_2} e^\beta \cdot \omega\;.
\]
If, moreover, $\pi_1\colon Y\to X_1$ is an oriented fiber bundle with compact closed oriented fibers, then we also have a Fourier-Mukai transform in the inverse direction, with kernel $-\beta$. Notice that by evident degree reasons the transforms $\Phi_\beta$ and $\Phi_{-\beta}$ are not inverses of one another. 
A particular way of obtaining a span of oriented fiber bundles $X_1\leftarrow Y\to X_2$ with compact closed oriented fibers is to consider a single oriented fiber bundle $Y\to Z$ with compact closed oriented fiber $F_1\times F_2$. Then the manifolds $X_1$ and $X_2$ are given by the total spaces of the $F_2$-fiber bundle and $F_1$-fiber bundles on $Z$, respectively, associated with the two factors of $F_1\times F_2$ together with the canonical projections. In particular, an oriented 2-torus bundle $Y\to Z$ produces this way a span $X_1\leftarrow Y\to X_2$ where both $\pi_i\colon Y\to X_i$ are $S^1$-bundles. It is precisely a configuration of this kind in which
 we will be interested.

\paragraph{From 1-form twists to 3-form twists.}
Assume now that $\alpha$ is a 3-form on $X$ instead of a 1-form. Then we can still define the operator $d_\alpha$ on differential forms as $d_\alpha\omega=d\omega+\alpha\wedge \omega$, but this will no more be a homogeneous degree 1 operator. We can heal this by adding a formal variable $u$ with $\deg(u)=2$ and with $du=0$, and define the degree 1 operator 
\[
d_\alpha\colon \Omega^\bullet(X)[[u^{-1},u]]\longrightarrow  \Omega^\bullet(X)[[u^{-1},u]]
\]
as the $\mathbb{R}[[u^{-1},u]]$-linear extension of 
\[
d_\alpha\omega = d\omega+u^{-1}\alpha\wedge \omega.
\]
Doing so, the above discussion verbatim applies, with the de Rham complex $\Omega^\bullet(X)$ replaced by the periodic de Rham complex  $\Omega^{{\bullet}}(X)[[u^{-1},u]]$. In particular, if we have a span $X_1\leftarrow Y\to X_2$ of oriented $S^1$-bundles and if $\alpha_i$ are 3-forms on $X_i$ such that $\pi_1^*\alpha_1-\pi_2^*\alpha_2=d\beta$ for some 2-form $\beta$ on $Y$, then we have Fourier-Mukai transforms
\begin{align*}
\Phi_\beta&\colon H_{\mathrm{dR};\alpha_1}^{{\bullet}}(X_1)[[u^{-1},u]] 
\longrightarrow
 H_{\mathrm{dR};\alpha_2}^{{\bullet}-{1}}(X_2)[[u^{-1},u]]\;,
 \\
\Phi_{-\beta}&\colon H_{\mathrm{dR};\alpha_2}^{{\bullet}}(X_2)[[u^{-1},u]] 
\longrightarrow
 H_{\mathrm{dR};\alpha_1}^{{\bullet}-{1}}(X_1)[[u^{-1},u]]\;.
\end{align*}
Having introduced the variable $u$, our cohomology is now endowed with a natural shift, given by the multiplication by $u$, and we may wonder whether the Fourier-Mukai transforms
$\Phi_\beta$ and $\Phi_{-\beta}$ may be inverses to one another up to shift. As we are going to see, this is precisely what happens in rational T-duality configurations.

\medskip
The above construction actually works for any closed differential form of odd degree, so there is apparently no point in considering 3-forms rather than 1-forms or 5-forms (see e.g. \cite{tcu}).
  There is, however, an important geometrical reason to focus on degree 3 forms: when the coefficients are taken in a characteristic zero field, periodic de Rham cohomology is isomorphic (via the Chern character) to $K$-theory. Under this isomorphism, $K$-theory twists (which are topologically given by principal $U(1)$-gerbes and so are classified by maps to $B^2U(1)\simeq K(\mathbb{Z},3)$) precisely become closed 3-forms. In other words, for $\alpha_1$ and $\alpha_2$ closed 3-forms as above, the Fourier-Mukai transform $\Phi_\beta$ is to be thought as a morphism (see \cite{BEM})
\[
\Phi_\beta\colon K_{\mathcal{G}_1}^{\bullet}(X_1)\otimes \mathbb{R} 
\longrightarrow
 K_{\mathcal{G}_2}^{\bullet-1}(X_2)\otimes \mathbb{R}\;.
\] 
where $\mathcal{G}_1$ and $\mathcal{G}_2$ are the twisting gerbes
(see \cite{BCMMS}). This is indeed the rationalization, with real coefficients, of a topological Fourier-Mukai transform
\[
\Phi_\beta\colon K_{\mathcal{G}_1}^{\bullet}(X_1)
\longrightarrow
 K_{\mathcal{G}_2}^{\bullet-1}(X_2)\;.
\]
\medskip
A particular situation we will be interested in is the case when the span  $X_1\leftarrow Y\to X_2$ of oriented $S^1$-bundles is induced by a 2-torus bundle $Y\to Z$, and so by a classifying map $Z\to B(U(1)\times U(1))\cong BU(1)\times BU(1)$. More specifically, we will also require that the canonical $U(1)$-2-gerbe associated with the torus bundle $Y\to Z$ is trivialized, i.e., we will be considering what is known as a topological T-duality configuration 
\cite{BunkeSchick05}. We will be investigating these from the point of view of rational homotopy theory, realizing the Fourier-Mukai transform as a morphism in twisted $L_\infty$-algebra cohomology and proving that a pair of $L_\infty$-algebras in a rational T-duality configuration comes equipped with a canonical Fourier-Mukai transform which turns out to be an isomorphism.

\subsection{Basics of rational homotopy theory}
\label{Sec-Basics}
The idea at the heart of rational homotopy theory is that, up to torsion, all of the homotopy type of a connected and simply connceted space\footnote{The theory can be extended to a \emph{simple space}, i.e., a connected topological space that has a homotopy type of a CW complex and whose fundamental group is abelian and acts trivially on the homotopy and homology of the universal covering space. A classical example is $S^1$, which we are actually going to meet several times in this note.} with finite rank cohomology groups is encoded in its de Rham algebra with coefficients in a characteristic zero field, as a differential graded commutative algebra, up to homotopy \cite{Quillen69} \cite{Sullivan77}. Moreover, since one has the freedom to replace the de Rham algebra with any homotopy equivalent DGCA, one sees that up to torsion the homotopy type of a simple space $X$ is encoded into its so called minimal model or Sullivan algebra: a DGCA $A_X$ equipped with a quasi-isomorphism of differential graded commutative algebras $A_X \to \Omega^\bullet(X)$, which is semi-free, i.e., which is a free graded commutative algebra when one forgets the differential, 
whose degree 1 component $A_X^1$ is zero,
and such that the differential is decomposable, i.e., it has no linear component. In other words, $A_X$ is a DGCA of the form $(\Exterior^\bullet \mathfrak{l}X^*,d)=(\mathrm{Sym}^\bullet(\mathfrak{l}X[1]^*),d)$ for a suitable graded vector space $\mathfrak{l}X$ concentrated in strictly negative degrees (and finitely dimensional in each degree)  and a suitable degree 1 differential $d$ with $d(\mathfrak{l}X^*)\subseteq \Exterior^{\geq 2}\mathfrak{l}X^*$. Here $\mathfrak{l}X^*$ denotes the graded linear dual of $\mathfrak{l}X$, and the degree shift in the definition of $\Exterior^\bullet$ is there in order to match the degree coming from geometry: the de Rham algebra is generated by 1-forms, which are in degree 1. 

\medskip
The semifreenes property together with the vanishing of $A_X^1$, the datum of the quasi-isomorphism to the de 
Rham algebra and the decomposability of the differential uniquely characterize the minimal model up to isomorphism and the quasi-isomorphism to the de Rham algebra up to homotopy, so that one can talk of \emph{the} minimal model of a space $X$. The pair $(\Exterior^\bullet \mathfrak{l}X^*,d)$ is what is called a \emph{minimal $L_\infty$-algebra} structure on $\mathfrak{l}X$ in the theory of $L_\infty$-algebras. Equivalently, one says that the DGCA $(\Exterior^\bullet \mathfrak{l}X^*,d)$ is the Chevalley-Eilenberg algebra of the $L_\infty$-algebra $\mathfrak{l}X$ (omitting the $L_\infty$ brackets of $\mathfrak{l}X$ from the notation), and writes 

\[
(A_X,d_X)\cong (\mathrm{CE}(\mathfrak{l}X),d_X)   
\]
as the defining equation of the $L_\infty$-algebra $\mathfrak{l}X$. 
We say that the $L_\infty$-algebra $\mathfrak{l}X$ is the rational approximation of $X$. Geometrically, it can be thought of as the tangent $L_\infty$-algebra to the $\infty$-group given by the based loop space of $X$ (as $X$ is connceted and simply connected, the choice of a basepoint is irrelevant). A smooth map $f\colon Y\to X$ is faithfully encoded into the DGCA morphism $f^*\colon \Omega^\bullet(X)\to \Omega^\bullet(Y)$, so that the rational approximation of $f$ is encoded into a DGCA morphism, which we will continue to denote $f^*$,
\[
f^*\colon A_Y\longrightarrow A_X.
\] 
In turn, by definition, this is a morphism of $L_\infty$-algebras $\mathfrak{l}f\colon \mathfrak{l}X\to \mathfrak{l}Y$. Finally, up to homotopy, every $L_\infty$-algebra is equivalent to a minimal one: this is the dual statement of the fact that every (well behaved) DGCA is homotopy equivalent to a minimal DGCA. Therefore we get the fundamental insight of rational homotopy theory: \emph{the category of simple homotopy types over a characteristic zero field $\mathbb{K}$ is (equivalent to) the homotopy category of $L_\infty$-algebras over $\mathbb{K}$.}

\begin{remark}
For non-simply connected simple spaces, one drops the condition $A_X^1=0$ and replaces it with the following nilpotency condition: one requires $\mathfrak{l}X^*$ to be filtered by an increasing series 
\[
\mathfrak{l}X^*(0)\subseteq \mathfrak{l}X^*(1)\subseteq \mathfrak{l}X^*(2)\subseteq \cdots\subseteq \mathfrak{l}X^*
\]
of graded subspaces with 
 $d(\mathfrak{l}X^*(0))=0$ and $d(\mathfrak{l}X^*(n))\subseteq \Exterior^{\geq 2}\mathfrak{l}X^*(n-1)$ for every $n\geq 1$. If $\mathfrak{l}X^*$ is finite-dimesnsional, this corresponds to requiring that $\mathfrak{l}X^*$ is a nilpotent $L_\infty$-algebra.
  When $A_X^1=0$ one has in particular that the degree 1 component of $\mathfrak{l}X[1]^*$ vanishes, and so the degree filtration on $\mathfrak{l}X^*$ automatically satisfies the nilpotency condition.
\end{remark}

The above description of rational homotopy theory may have erroneously suggested it is a quite abstract construction. So let us make a few examples to make it concrete.
\begin{example}[The Sullivan model of $BU(1)$]
 Consider the classifying space $BU(1)$. Its real cohomology is $H^\bullet(BU(1);\mathbb{R})\cong \mathbb{R}[x_2]$, where $x_2$ is a degree 2 element, the universal first Chern class. As $H^\bullet(BU(1);\mathbb{R})$ is a free polynomial algebra, we can think of it as a semifree DGCA with trivial differential. Moreover, choosing a de Rham representative for the first Chern Class defines a quasi-isomorphism
\[
(\mathbb{R}[x_2],0) \longrightarrow (\Omega^\bullet(BU(1)),d)
\]
exhibiting $(\mathbb{R}[x_2],0)$ as the Sullivan model of $BU(1)$. The equation
\[
(\mathbb{R}[x_2],0)\cong (\mathrm{CE}(\mathfrak{l}BU(1)),d_{BU(1)})  
\]
then characterizes $\mathfrak{l}BU(1))$ as the $L_\infty$-algebra consisting of the cochain complex $\mathbb{R}[1]$ consisting of the vector space $\mathbb{R}$ in degree -1 and zero in all other degrees (with zero differential). We will denote this $L_\infty$-algebra by the symbol $b\mathfrak{u}_1$. A principal $U(1)$-bundle $P\to X$ is classified by a map $X\to BU(1)$. The rational approximation of this map is an $L_\infty$-morphism
\[
\mathfrak{l}X \longrightarrow b\mathfrak{u}_1.
\]
Equivalently, by definition, this is a DGCA morphism
\[
(\mathbb{R}[x_2],0)\longrightarrow (A_X,d_X),
\]
i.e., it is a degree 2 closed element in $A_X$. By pushing it forward along the quasi-isomorphism $(A_X,d_X)\xrightarrow{\sim} (\Omega^\bullet(X),d)$ we get a closed 2-form $\omega_2$ on $X$ associated to the principal $U(1)$-bundle $P\to X$. Since the quasi-isomorphism $(A_X,d_X)\xrightarrow{\sim} (\Omega^\bullet(X),d)$ is only unique up to homotopy, the 2-form $\omega_2$ is only well defined up to an exact term so that it is the cohomology class $[\omega_2]$ to be actually canonically associated with $P\to X$. No surprise, $[\omega_2]$ is the image in de Rham cohomology of the first Chern class of $P\to X$. 
\end{example}

\begin{example}[compact abelian Lie groups]Another classical example is the following. Given a compact Lie group $G$, then the inclusion $\Omega^\bullet(G)^G\hookrightarrow \Omega^\bullet$ of $G$-invariant differential forms on $G$ into the de Rham complex of $G$ is a quasi-isomorphism. As a $G$-invariant form is completely and freely determined by its restriction at the identity element of $G$, we see that as a graded vector space $\Omega^\bullet(G)^G\cong \Exterior^\bullet\mathfrak{g}^*$, where $\mathfrak{g}$ denotes the Lie algebra of $G$. The de Rham differential on $\Omega^\bullet(G)^G$ corresponds to the Chevalley-Eilenberg differential on $\Exterior^\bullet\mathfrak{g}^*$, i.e., to the differential computing the Lie algebra cohomology of $\mathfrak{g}$ with coefficients in $\mathbb{R}$ as a trivial $\mathfrak{g}$-module. From this we see that a semifree model for $G$ is $\mathrm{CE}(\mathfrak{g})$. However, $\mathrm{CE}(\mathfrak{g})$ is \emph{not} a Sullivan model for $G$, unless $\mathfrak{g}$ is nilpotent. This happens in particular for compact abelian Lie groups, so that, for instance $\mathrm{CE}(\mathfrak{u}_1)$ is indeed the Sullivan model of $U(1)$. 

\end{example}

\begin{example} 
[The Sullivan models of spheres]
The real cohomology ring of the $n$-sphere $S^n$ is extremely simple: we have
\[
H^\bullet(S^n;\mathbb{R})\simeq \begin{cases}
\mathbb{R}[t_{n}] & \text{if $n$ is odd}
\\
\mathbb{R}[t_{n}]/({t_n}^2) & \text{if $n$ is even}
\end{cases}
\]
as graded commutative rings, where $t_n$ is a variable in degree $n$. Notice that the difference between the odd and the even case is only apparent: in the odd case ${t_n}^2=0$ due to the graded commutativity of the product. However, we preferred to divide the two cases to stress that in the odd case, the rational cohomology of $S^n$ is a free graded polynomial algebra, and so it essentially coincides with its own Sullivan model, we only need to add a trivial differential to the picture:
\[
{\rm CE}(\mathfrak{l}S^{2k+1})=\big(\mathbb{R}[x_{2k+1}];~dx_{2k+1}=0 \big).
\]
Namely, if $\omega_{2k+1}$ is a volume form for $S^{2k+1}$, the map $x_{2k+1}\mapsto \omega_{2k+1}$ defines a morphism of differential graded commutative algebras
\[
\big(\mathbb{R}[x_{2k+1}];~dx_{2k+1}=0 \big) 
\longrightarrow
\big(\Omega^\bullet(S^{2k+1};\mathbb{R});~ d_{\mathrm{dR}}\big)
\]
which is immediately seen to be a quasi-isomorphism, i.e., inducing an isomorphism in cohomology.
For even $n=2k$ we have to cure the constraint ${t_{2k}}^2=0$. This is done by lifting the cohomology relation ${t_{2k}}^2=0$ to the equation ${x_{2k}}\wedge x_{2k}=d x_{4k-1}$. It is then easy to see that, if we consider the free polynomial algebra $\mathbb{R}[x_{2k}, x_{4k-1}]$ and introduce on it the differential $d$ by the rule $dx_{2k}=0$ and $dx_{4k-1}={x_{2k}}^2$ then we see that $\big(\mathbb{R}[x_{2k}, x_{4k-1}];~dx_{2k}=0,\, dx_{4k-1}={x_{2k}}\wedge x_{2k} \big)$ is a differential graded commutative algebra and that
\begin{align*}
\big(\mathbb{R}[x_{2k}, x_{4k-1}];~dx_{2k}=0,\, dx_{4k-1}={x_{2k}}\wedge x_{2k} \big)&\longrightarrow \big(\Omega^\bullet(S^{2k};\mathbb{R});~ d_{\mathrm{dR}}\big)\\
x_{2k}&\longmapsto \omega_{2k}\\
x_{4k-1}&\longmapsto 0
\end{align*}
is a quasi-isomorphism of DGCAs. Moreover, $\mathbb{R}[x_{2k}, x_{4k-1}]^1=0$ and the differential is decomposable. In other words,
\[
{\rm CE}(\mathfrak{l}S^{2k})=\big(\mathbb{R}[x_{2k}, x_{4k-1}];~dx_{2k}=0,\, dx_{4k-1}={x_{2k}}\wedge x_{2k} \big).
\]
\end{example}
\vskip .8 cm

Given the identification between simple homotopy types and $L_\infty$-algebras mentioned above, from now on we will mostly work directly with $L_\infty$-algebras, with no reference to the space of which they can be a rationalization. Therefore, a span $X_1\leftarrow Y \to X_2$ as in the discussion of Fourier-Mukai transforms in twisted de Rham cohomology will become a span
\[
\xymatrix@R=1.3em{
& \mathfrak{h}\ar[dl]_{\pi_1}\ar[dr]^{\pi_2}&\\
\mathfrak{g}_1&&\mathfrak{g}_2
}
\]
of $L_\infty$-algebras. As we want that the $\pi_i$'s represent the (rationalization of) $S^1$-bundles our next step is the characterization of those $L_\infty$-morphism that correspond to principal $U(1)$-bundles.

\subsection{Central extensions of $L_\infty$-algebras}\label{section.central}
A principal $U(1)$-bundle over a smooth manifold $X$ is encoded up to homotopy into a map $f\colon X\to BU(1)$ from $X$ to the classifying space $U(1)$. The total space $P$ as well as the projection $P\to X$ are recovered by $f$ by taking its homotopy fiber, i.e., by considering the homotopy pullback
\[
\xymatrix@R=1.3em
{P\ar[rr]\ar[d]&&{*}\ar[d]\\
X\ar[rr]^-f && BU(1)\;.
}
\]
As rationalization commutes with homotopy pullbacks, the rational approximation 
of the above diagram is
\[
\xymatrix@R=1.3em{
\mathfrak{l}P\ar[rr]\ar[d]&&{0}\ar[d]\\
\mathfrak{l}X\ar[rr]^{\mathfrak{l}f}&&b\mathfrak{u}_1
\;.}
\]
Dually, this means that we have a homotopy pushout of DGCAs
\[
\xymatrix@=1.5em{
(\mathbb{R}[x_2],0)\ar[rr]\ar[d]_{f^*}&&
(\mathbb{R},0)\ar[d]\\
(A_X,d_X)\ar[rr]&&(A_P,d_P)
\;.
}
\]
This is easily computed. All we have to do is to replace the DCGA morphism 
$\mathbb{R}[x_2]\to \mathbb{R}$ with an equivalent cofibration. The 
easiest way of doing this is to factor $\mathbb{R}[x_2]\to \mathbb{R}$ as 
\[
\xymatrix{
(\mathbb{R}[x_2],0)\; \ar@{^{(}->}[r] &
 (\mathbb{R}[y_1,x_2], dy_1=x_2) \ar[r]^-{\sim} & \mathbb{R}
}
\]
Then $A_P$ is computed as an ordinary pushout
\[
\xymatrix@=1.5em{
(\mathbb{R}[x_2],0)\ar[rr]\ar[d]_{f^*}&&
(\mathbb{R}[y_1,x_2], dy_1=x_2)\ar[d]\\
(A_X,d_X)\ar[rr] && (A_P,d_P)
\;,}
\]
 i.e.,
\[
(A_P,d_P)=(A_X[y_1], d_P\omega=d_X\omega \;\text{ for }\; \omega\in A_X, \, d_Py_1=f^*x_2).
\]
This immediately generalizes to the case of an arbitrary morphism $f\colon \mathfrak{g}\to b\mathfrak{u}_1$. The homotopy fiber of $f$ will be the $L_\infty$-algebra $\hat{\mathfrak{g}}$  characterized by
\[
\mathrm{CE}(\hat{\mathfrak{g}})=\mathrm{CE}(\mathfrak{g})[y_1],
\]
where $y_1$ ia a variable in degree 1 and where the differential in $\mathrm{CE}(\hat{\mathfrak{g}})$ extends that in $\mathrm{CE}(\mathfrak{g})$ by the rule $d_{\hat{\mathfrak{g}}}y_1=f^*(x_2)$.
\begin{example}
If $\mathfrak{g}$ is a Lie algebra (over $\mathbb{R}$), then an $L_\infty$-morphism $f\colon \mathfrak{g}\to b\mathfrak{u}_1$ is precisely a Lie algebra 2-cocycle on $\mathfrak{g}$ with values in $\mathbb{R}$. The $L_\infty$-algebra $\hat{\mathfrak{g}}$ is again a Lie algebra in this case, and it is the central extension of $\mathfrak{g}$ by $\mathbb{R}$ classified by the 2-cocycle $f$.
\end{example}
The above construction admits an immediate generalization. Instead of $b\mathfrak{u}_1$ we can consider the $L_\infty$-algebra $b^n\mathfrak{u}_1$ given by the cochain complex $\mathbb{R}[n]$ consisting of $\mathbb{R}$ in degree $-n$ and zero in all other degrees. The corresponding Chevalley-Eilenberg algebra is 
\[
(\mathrm{CE}(b^n\mathfrak{u}_1),d_{b^n\mathfrak{u}_1})=(\mathbb{R}[x_{n+1}],0),
\]
where $x_{n+1}$ is in degree $n+1$. One sees that $b^n\mathfrak{u}_1$ is a rational model (with coefficients in $\mathbb{R}$) for the classifying space $B^nU(1)$ of principal $U(1)$-$n$ bundles (or principal $U(1)$-$(n-1)$-gerbes), which is a $K(\mathbb{Z},n+1)$. If $\mathfrak{g}$ is a Lie algebra, then an $L_\infty$-morphism $\mathfrak{g}\to b^n\mathfrak{u}_1$ is precisely a Lie algebra $(n+1)$-cocycle on $\mathfrak{g}$ with coefficients in $\mathbb{R}$. More generally, an $L_\infty$-morphism  $\mathfrak{g}\to b^n\mathfrak{u}_1$ with $\mathfrak{g}$ an $L_\infty$-algebra will also be called an $(n+1)$-cocycle. The dual picture makes this terminology transparent: an $(n+1)$-cocycle on $\mathfrak{g}$ is a DGCA morphism
\[
(\mathbb{R}[x_{n+1}],0)\longrightarrow (\mathrm{CE}(\mathfrak{g}),d_{\mathfrak{g}})
\]
so it is precisely a closed degree $n+1$ element in $\mathrm{CE}(\mathfrak{g})$. The description of homotopy fibers of 2-cocycles immediately generalizes to higher cocycles: the homotopy fiber $\hat{\mathfrak{g}}$ of an $(n+1)$-cocycle  $\mathfrak{g}\to b^n\mathfrak{u}_1$ is characterized by
\[
\mathrm{CE}(\hat{\mathfrak{g}})=\mathrm{CE}(\mathfrak{g})[y_n],
\]
where $y_n$ is a variable in degree $n$ and where the differential in $\mathrm{CE}(\hat{\mathfrak{g}})$ extends that in $\mathrm{CE}(\mathfrak{g})$ by the rule $d_{\hat{\mathfrak{g}}}y_1=f^*(x_{n+1})$. By analogy with the case of 
2-cocycles on Lie algebras, one calls $\hat{\mathfrak{g}}$ a higher central extension of $\mathfrak{g}$. Geometrically, 
$\hat{\mathfrak{g}}$ is to be thought as the total space of a rational $U(1)$-$n$-bundle over $\mathfrak{g}$.

\subsection{Twisted $L_\infty$-algebra cohomology}
As we explained in Section \ref{Sec-Basics}, a (finite dimensional in each degree) $L_\infty$-algebra $\mathfrak{g}$ is encoded into its Chevalley-Eilenberg algebra $(\mathrm{CE}(\mathfrak{g}),d_{\mathfrak{g}})$. As this is a
DGCA, we can consider its cohomology which, by definition, is the $L_\infty$-algebra cohomology of $\mathfrak{g}$
\[
H^\bullet_{L_\infty}(\mathfrak{g};\mathbb{R})=H^\bullet\left( \mathrm{CE}(\mathfrak{g}),d_{\mathfrak{g}}\right). 
\]
When $\mathfrak{g}$ is a Lie algebra this reproduces the Lie algebra cohomology of $\mathfrak{g}$.
If $\mathfrak{g}$ is the $L_\infty$-algebra representing the rational 
homotopy type of a simple space $X$, then the $L_\infty$-algebra cohomology of $\mathfrak{g}$ computes the de Rham cohomology of $X$. That is,
\[
H^\bullet_{L_\infty}(\mathfrak{l}X;\mathbb{R})=H^\bullet\left( \mathrm{CE}(\mathfrak{l}X),d_X\right)=H^\bullet\left( A_X,d_X)\right)\cong H^\bullet\left( \Omega^\bullet(X),d)\right)=H^\bullet_{\mathrm{dR}}(X). 
\]
This is more generally true if instead of the Sullivan model $\mathrm{CE}(\mathfrak{l}X)$ one considers an arbitrary semifree model $\mathrm{CE}(\mathfrak{g}_X)$.
\begin{example}
If $\mathfrak{g}$ is the Lie algebra of a compact Lie group $G$, then one recovers the classical statement that the Lie algebra cohomology of $\mathfrak{g}$ computes the de Rham cohomology of $G$:
\[
H^\bullet_{\mathrm{Lie}}(\mathfrak{g};\mathbb{R})\cong H^\bullet_{\mathrm{dR}}(G).
\]
This has actually been one of the motivating examples in the definition of Lie algebra cohomology. 
\end{example}
Exactly as we twisted de Rham cohomology in Section \ref{Sec-twdR},
we can twist $L_\infty$-algebra cohomology: if $a$ is a degree 3 cocycle on $\mathfrak{g}$ then we can consider the degree 1 differential $d_{\mathfrak{g};a}\colon x\mapsto d_\mathfrak{g}x+u^{-1}a\,x$ on the algebra of Laurent series in the variable $u$ with coefficients in the Chevalley-Eilenberg algebra of $\mathfrak{g}$ and define
\[
H^{\bullet}_{L_\infty;a}(\mathfrak{g};\mathbb{R})[[u^{-1},u]]=H^{\bullet}\left( \mathrm{CE}(\mathfrak{g})[[u^{-1},u]],d_{\mathfrak{g};a}\right). 
\]
As in the de Rham case, if $a_1$ and $a_2$ are cohomologous 3-cocycles with $a_1-a_2=db$ then $e^{u^{-1}b}$ is a cochain complexes isomorphism between $(\mathrm{CE}(\mathfrak{g})[[u^{-1},u]],d_{\mathfrak{g};a_1})$ and $(\mathrm{CE}(\mathfrak{g})[[u^{-1},u]],d_{\mathfrak{g};a_2})$ and so induces an isomorphism
\[
e^{u^{-1}b}\colon H^{{\bullet}}_{L_\infty;a_1}(\mathfrak{g};\mathbb{R})[[u^{-1},u]]\overset{\sim}{\longrightarrow} H^{{\bullet}}_{L_\infty;a_2}(\mathfrak{g};\mathbb{R})[[u^{-1},u]]\;.
\]
If $f\colon \mathfrak{h}\to \mathfrak{g}$ is an $L_\infty$ morphism, then by definition $f$ is a DGCA morphism $f^*\colon \mathrm{CE}(\mathfrak{g})\to \mathrm{CE}(\mathfrak{h})$ so that $f^*a$ is a 3-cocycle on $\mathfrak{h}$ for any 3-cocycle $a$ on $\mathfrak{g}$, and $f^*$ is a morphism of cochain complexes between $(\mathrm{CE}(\mathfrak{g})[[u^{-1},u]],d_{\mathfrak{g};a})$ and $(\mathrm{CE}(\mathfrak{h})[[u^{-1},u]],d_{\mathfrak{h};f^*a})$, thus inducing a morphism between the twisted cohomologies
\[
f^*\colon H^{{\bullet}}_{L_\infty;a}(\mathfrak{g};\mathbb{R})[[u^{-1},u]]
\longrightarrow H^{{\bullet}}_{L_\infty;f^*a}(\mathfrak{h};\mathbb{R})[[u^{-1},u]].
\]
We, therefore, see that in order to define Fourier-Mukai transforms at the level of twisted $L_\infty$-algebra cohomology the only ingredient we miss is a pushforward morphism
\[
\pi_*\colon (\mathrm{CE}(\hat{\mathfrak{g}}),d_{\hat{\mathfrak{g}}})
\longrightarrow
\big(\mathrm{CE}(\mathfrak{g})[-1],d_{\mathfrak{g}}[-1]\big)
\]
for any central extension $\pi\colon \hat{\mathfrak{g}}\to \mathfrak{g}$ induced by a 2-cocycle $\mathfrak{g}\to b\mathfrak{u}_1$, which is a morphism of cochain complexes and which satisfies the projection formula identity. We are going to exhibit such a morphism in the next section.

\subsection{Fiber integration along $U(1)$-bundles in rational homotopy theory}
\label{fiber-integration}
Let $P\to X$ be a principal $U(1)$-bundle. Since $U(1)$ is a compact Lie group, every differential form on $P$ can be averaged so to become invariant under the $U(1)$-action on $P$. Moreover, taking average is a homotopy inverse to the inclusion of $U(1)$-invariant forms into all forms on $P$ so that
\[
\xymatrix{
\Omega^\bullet(P)^{U(1)}\; \ar@{^{(}->}[r] & \Omega^\bullet(P)
}
\]
is a quasi-isomorphism of DGCAs. The DGCA $\Omega^\bullet(P)^{U(1)}$ has a very simple description in terms of the DGCA $\Omega^\bullet(X)$. Namely, identifying $\Omega^\bullet(X)$ with its image in $\Omega^\bullet(P)$ via $\pi^*$ one sees that $\Omega^\bullet(X)$ is actually a subalgebra of $\Omega^\bullet(P)^{U(1)}$. The subalgebra $\Omega^\bullet(X)$ however does not exhaust all of the $U(1)$-invariant forms on $P$: those forms that restrict to a scalar multiple of the volume form on the fibers (for some choice of a $U(1)$-invariant metric on $P$) are left out. Picking one such a form $\omega_1$ is equivalent to the datum of a $U(1)$-connection $\nabla$ on $P$ and
\[
(\Omega^\bullet(P)^{U(1)},d)=(\Omega^\bullet(X)[\omega_1], d\omega_1=F_\nabla),
\]
where $F_\nabla$ is the curvature of $\nabla$, so that we have a quasi-isomorphism of DGCAs 
\[
(\Omega^\bullet(X)[\omega_1], d\omega_1=F_\nabla)\xrightarrow{\;\;\sim\;\;}(\Omega^\bullet(P),d).
\]
This is the geometric counterpart of the isomorphism
\[
(\mathrm{CE}(\hat{\mathfrak{g}}),d_{\hat{\mathfrak{g}}})=(\mathrm{CE}(\mathfrak{g})[y_1], d_{\hat{\mathfrak{g}}}y_1=f^*x_2)
\]
we met in Section 3, so that we see that the degree 1 element $y_1$ in the Chevalley-Eilenberg of the central extension $\hat{\mathfrak{g}}$ does indeed represent a vertical volume form. The fiber integration $\pi_*\colon (\Omega^\bullet(P),d) \to (\Omega^\bullet(X)[-1],d[-1])$, restricted to $U(1)$-invariant forms reads
\begin{align*}
\pi_*\colon (\Omega^\bullet(X)[\omega_1], d\omega_1=F_\nabla) &
\longrightarrow (\Omega^\bullet(X)[-1],d[-1])\\
\alpha+\omega_1\wedge \beta &\longmapsto \beta,
\end{align*}
so it is natural to define the fiber integration morphism $\pi_*$ associated with the central extension $\pi\colon \hat{\mathfrak{g}}\to \mathfrak{g}$ determined by the 2-cocycle $f\colon \mathfrak{g}\to \mathfrak{u}_1$ as
\begin{align*}
\pi_*\colon (\mathrm{CE}(\mathfrak{g})[y_1], d_{\hat{\mathfrak{g}}}y_1=f^*x_2) &
\longrightarrow (\mathrm{CE}(\mathfrak{g})[-1],d_{\mathfrak{g}}[-1])\\
a+y_1\,b &\longmapsto b,
\end{align*}
It is immediate to see that $\pi_*$ is indeed a morphism of chain complexes:
\[
d_{\mathfrak{g}}[-1](\pi_*(a+y_1\,b))=-d_{\mathfrak{g}}b=   \pi_*(d_{\mathfrak{g}}a+ (f^*x_2)\, b-y_1\,d_{\mathfrak{g}}b )=\pi_*(d_{\hat{\mathfrak{g}}}(a+y_1\,b)).
\]
Next, let us show that the projection formula holds. Since the morphism $\pi^*\colon (\mathrm{CE}(\mathfrak{g}), d_{\mathfrak{g}})\to (\mathrm{CE}(\hat{\mathfrak{g}}),d_{\hat{\mathfrak{g}}})$ is the inclusion of $\mathrm{CE}(\mathfrak{g})$ into $\mathrm{CE}(\mathfrak{g})[y_1]$, we find:
\[
\pi_*((\pi^*a)\, (b+y_1\, c)) 
=\pi_*(a\, b+ (-1)^ay_1\, ac))
=(-1)^a ac
=(-1)^aa\, \pi_*(b+y_1\, c)
\]
for every $b,c\in \mathrm{CE}(\mathfrak{g})$, i.e.,
\begin{equation}\label{eq.projection}
\pi_*((\pi^*a)\, \omega )=(-1)^aa\, \pi_*\omega,
\end{equation}
for every $\omega\in \mathrm{CE}(\hat{\mathfrak{g}})$.
Summing up, we have reproduced at the $L_\infty$-algebra/rational homotopy theory level all of the ingredients we needed to define Fourier-Mukai transforms. That is, give a span $\mathfrak{g}_1\xleftarrow{\pi_1} \mathfrak{h}\xrightarrow{\pi_2} \mathfrak{g}_2$ of central extensions (by the abelian Lie algebra $\mathbb{R}$) of $L_\infty$-algebras, and given a triple $(a_1,a_2,b)$ consisting
 of 3-cocycles $a_i$ on $\mathfrak{g}_i$ and of a degree 2 element $b$ in $\mathrm{CE}(\mathfrak{h})$ such that $d_\mathfrak{h}b=\pi_1^*a_1-\pi_2^*a_2$ we have Fourier-Mukai transforms
\begin{align*}
\Phi_b&\colon H_{L_\infty;a_1}^{{\bullet}}(\mathfrak{g}_1;\mathbb{R})[[u^{-1},u]] 
\longrightarrow H_{L_\infty;a_2}^{{\bullet}-{1}}(\mathfrak{g}_2;\mathbb{R})[[u^{-1},u]]\\
\Phi_{-b}&\colon H_{L_\infty;a_2}^{{\bullet}}(\mathfrak{g}_2;\mathbb{R})[[u^{-1},u]] 
\longrightarrow H_{L_\infty;a_1}^{{\bullet}-{1}}(\mathfrak{g}_1;\mathbb{R})[[u^{-1},u]]
\end{align*}
given by the images in cohomology of the morphisms of complexes
\[
\omega\longmapsto \pi_{2*}(e^{u^{-1}b_2}\pi_1^*\omega)\qquad \text{ and } \qquad\omega\longmapsto \pi_{1*}(e^{-u^{-1}b_2}\pi_2^*\omega),
\]
respectively.  

\subsection{The hofiber/cyclification adjunction and cyclification of $L_\infty$-algebras}
\label{cyc-l-infty}
We are going to see how to produce a quintuple $(\pi_1,\pi_2,a_1,a_2,b)$ inducing a Fourier-Mukai transform in Section \ref{rht-tfold}. But first let us spend a few more words on the geometric properties of the pushforward morphism $\pi_*$. As 
$\pi_*\colon (\mathrm{CE}(\hat{\mathfrak{g}}), d_{\hat{\mathfrak{g}}}) \to (\mathrm{CE}(\mathfrak{g})[-1],d_{\mathfrak{g}}[-1])$ is a morphism of cochain complexes, it in particular maps degree $n+1$ cocycles in $\mathrm{CE}(\hat{\mathfrak{g}})$ to degree $n$ cocycles in $\mathrm{CE}(\mathfrak{g})$. But, if $\mathfrak{h}$ is any $L_\infty$-algebra, we have seen that a degree $k$ cocycle in $\mathrm{CE}(\mathfrak{h})$ is precisely an $L_\infty$-morphism $\mathfrak{h}\to b^{k-1}\mathfrak{u}_1$. Therefore we see that $\pi_*$ induces a morphism of sets
\[
\mathrm{Hom}_{L_\infty}(\hat{\mathfrak{g}},b^{n}\mathfrak{u}_1)
\longrightarrow
 \mathrm{Hom}_{L_\infty}({\mathfrak{g}},b^{n-1}\mathfrak{u}_1).
\]
This is actually part of a much larger picture, to see which we need a digression on free loop spaces. So let again $X$ be our smooth manifold and let $\pi\colon P\to X$ be a principal $U(1)$-bundle over $X$, and let $\varphi\colon P\to Y$ a map from $P$ to another smooth manifold $Y$. Let $\gamma\colon P\times U(1) \to Y$
be the composition
\[
P\times U(1) \longrightarrow P \xrightarrow{\;\varphi\;} Y
\]
where the first map is the right $U(1)$-action on $P$. By the multiplication by $S^1$/free loop space adjunction, $\gamma$ is, equivalently, a morphism from $P$ to the free loop space $\mathcal{L}Y$ of $Y$. More explicitly, a point $x\in P$ is mapped to the loop $\varphi_x\colon U(1)\to Y$ defined by $\varphi_x(e^{i\theta})=\varphi(x\cdot e^{i\theta})$. The map $\gamma\colon P\to \mathcal{L}Y$ is equivariant with respect to the right $U(1)$-action on $P$ and the right $U(1)$-action on $\mathcal{L}Y$ given by loop rotation: $\eta\cdot e^{i\theta}=\rho_\theta^*\eta$, where $\rho_\theta\colon U(1)\to U(1)$ is the rotation by angle $\theta$. Namely, one has
\[
((\varphi_x)\cdot e^{i\theta})(e^{i\theta_0}) =(\rho_\theta^*\varphi_x)(e^{i\theta_0})=\varphi_x(e^{i\theta}e^{i\theta_0})=\varphi((x\cdot e^{i\theta})\cdot e^{i\theta_0})
=\varphi_{x\cdot e^{i\theta}}(e^{i\theta_0}).
\]
Therefore, equivalently, $\gamma$ is a morphism between the homotopy quotients $P//U(1)$ and $\mathcal{L}Y//U(1)$ over $BU(1)$. Moreover, as $P$ is a principal $U(1)$-bundle over $X$, the homotopy quotient $P/\!/U(1)$ is equivalent to the ordinary quotient and so is equivalent to the base $X$, and the natural map $P/\!/U(1)\to BU(1)$ is identified with the morphism $f\colon X \to BU(1)$ classifying the principal bundle $P$. In other words, a morphism $\varphi\colon P\to Y$ is,
 equivalently, a morphism
\[
\xymatrix@R=1.4em{
X\ar[rr]\ar[rd]_f&& \mathcal{L}Y/\!/U(1)\ar[ld]\\
&BU(1)
}
\]
from $f$ to the canonical morphism $\mathcal{L}Y/\!/U(1)\to BU(1)$ in the overcategory of spaces over $BU(1)$. Writing $\mathrm{cyc}(Y)$ for the ``cyclification'' $\mathcal{L}Y/\!/U(1)$ and recalling that the total space $P$ is the homotopy fiber of the morphism $f\colon X\to BU(1)$, we see that the above discussion can be elegantly summarized by saying that cyclification is the right adjoint to homotopy fiber,
\[
\xymatrix@R=1.5em{
\mathrm{spaces}\ar@/^1.6pc/[rr]^{\mathrm{cyc}}&&\mathrm{spaces}/BU(1)\ar@/^1.6pc/[ll]^{\mathrm{hofib}}.
}
\]
\paragraph{Cyclification of $L_\infty$-algebras.}
The above topological construction immediately translates to the $L_\infty$-algebra setting, 
where we find an adjunction
\footnote{A more general statement and proof in $\infty$-toposes can be found in 
\cite{urs} (see Proposition 4.1).}
\[
\xymatrix{
L_\infty\mathrm{\text{-}algebras}\ar@/^1.6pc/[rr]^{\mathrm{cyc}}&&L_\infty\mathrm{\text{-}algebras}/b\mathfrak{u}_1\ar@/^1.6pc/[ll]^{\mathrm{hofib}}.
}
\]
We have already seen that the homotopy fiber functor from $L_\infty$-algebras over $b\mathfrak{u}_1$(i.e., $L_\infty$-algebras equipped with an $\mathbb{R}$-valued 2-cocycle) to $L_\infty$-algebras consists in forming the $\mathbb{R}$-central extension classified by the 2-cocycle. So we have now to complete the picture by describing the cyclification functor. As usual, we start from geometry, and consider an $L_\infty$-algebra $\mathfrak{l}X$ representing the rational homotopy type of a simple space $X$. If $X$ is 2-connected (so that its free loop space is surely simply connected and therefore simple) an $L_\infty$-algebra representing the rational homotopy type of the free loop space $\mathcal{L}X$ is easily deduced from the multiplication by $S^1$/free loop space adjunction. As a Sullivan model for $Y\times S^1$ is $A_{Y\times S^1}=A_Y\otimes A_{S^1}=A_Y[t_1]$ with $dt_1=0$, one sees that if $A_X=(\Exterior^\bullet \mathfrak{l}X^*,d_X)$, then
\[
A_{\mathcal{L}X}=(\Exterior^\bullet (\mathfrak{l}X^*\oplus s\mathfrak{l}X^*),d_{\mathcal{L}X})
\]
where $s\mathfrak{l}X^*=\mathfrak{l}X^*[1]$ is a shifted copy of $\mathfrak{l}X^*$, 
with $d_{\mathcal{L}X}\bigr\vert_{A_X}=d_X$ and $[d_{\mathcal{L}X},s]=0$, where $s\colon A_{\mathcal{L}X}\to A_{\mathcal{L}X}$ is the shift operator $s\colon \mathfrak{l}X^*\xrightarrow{\sim} (s\mathfrak{l}X^*)[-1]$ extended as a degree -1 differential. See \cite{VigueBurghelea} for details. This immediately suggests the following definition: for an arbitrary $L_\infty$-algebra $\mathfrak{g}$ we write $\mathfrak{L}\mathfrak{g}$ for the $L_\infty$-algebra defined by
\[
(\mathrm{CE}(\mathcal{L}\mathfrak{g}),d_{\mathfrak{L}\mathfrak{g}})=
\big(\Exterior^\bullet (\mathfrak{g}^*\oplus s\mathfrak{g}^*),d_{\mathfrak{L}\mathfrak{g}}\bigr\vert_{\mathrm{CE}(\mathfrak{g})}=d_{\mathfrak{g}}, [d_{\mathfrak{L}\mathfrak{g}},s]=0]\big).
\]
Deriving an $L_\infty$-algebra model for the cyclification $\mathrm{cyc}(X)$ is a bit more involved, ad has been worked out in \cite{VigueSullivan}. One finds
\[
A_{\mathrm{cyc}(X)}=
\big(\Exterior^\bullet (\mathfrak{l}X^*\oplus s\mathfrak{l}X^*\oplus b\mathfrak{u}_1^*),d_{\mathrm{cyc}(X)}\big)
=\big(\Exterior^\bullet (\mathfrak{l}X^*\oplus s\mathfrak{l}X^*)[x_2],d_{\mathrm{cyc}(X)}\big),
\]
where $x_2$ is a degree 2 closed variable and  $d_{\mathrm{cyc}(X)}$ acts on an element $a\in \mathfrak{l}X^*\oplus s\mathfrak{l}X^*$ as $
d_{\mathrm{cyc}X}a=d_{\mathfrak{L}\mathfrak{g}}a+x_2\wedge sa$.
From this one has the natural generalization to an arbitrary $L_\infty$-algebra $\mathfrak{g}$: its cyclification is the $L_\infty$-algebra $\mathrm{cyc}(\mathfrak{g})$ defined by
\[
\mathrm{CE}(\mathrm{cyc}(\mathfrak{g}))=
\big(\Exterior^\bullet (\mathfrak{g}\oplus s\mathfrak{g}\oplus b\mathfrak{u}_1)^*,d_{\mathrm{cyc}(\mathfrak{g})}\big)
=\big((\Exterior^\bullet (\mathfrak{g}\oplus s\mathfrak{g})^*)[x_2],d_{\mathrm{cyc}(\mathfrak{g})}\big),
\]
where $x_2$ is a degree 2 variable with $d_{\mathrm{cyc}(\mathfrak{g})}x_2=0$ and  $d_{\mathrm{cyc}(\mathfrak{g})}$ acts on an element $a\in \mathfrak{g}^*[-1]\oplus \mathfrak{g}^*$ as 
\[
d_{\mathrm{cyc}(\mathfrak{g})}a=d_{\mathfrak{L}\mathfrak{g}}a+x_2\wedge sa.
\]
Notice that there is a canonical inclusion of DGCAs $\mathbb{R}[x_2]\hookrightarrow \mathrm{CE}(\mathrm{cyc}(\mathfrak{g}))$, giving a canonical 2-cocycle $\mathrm{cyc}(\mathfrak{g})\to b\mathfrak{u}_1$. 
It is then not hard to see that, if $f\colon \mathfrak{g}\to b\mathfrak{u}_1$ is a 2-cocycle classifying a central extenson $\hat{\mathfrak{g}}$, then there is a natural bijection
\[
\mathrm{Hom}_{L_\infty}(\mathrm{hofib}(f),\mathfrak{h}) \cong \mathrm{Hom}_{L_\infty/b\mathfrak{u}_1}(\mathfrak{g},\mathrm{cyc}(\mathfrak{h})),
\]
for any $L_\infty$-algebra $\mathfrak{h}$, where on the right hand side with a little abuse of notation we have written the sources in places of the morphisms. Namely,  in the dual Chevalley-Eilenberg 
picture this amounts to a natural bijection
\[
\mathrm{Hom}_{\rm DGCA}(\mathrm{CE}(\mathfrak{h}), \mathrm{CE}(\hat{\mathfrak{g}})) \cong \mathrm{Hom}_{R[x_2]/{\rm DGCA}}(\mathrm{CE}(\mathrm{cyc}(\mathfrak{h})),\mathrm{CE}(\mathfrak{g})).
\]
As $\mathrm{CE}(\mathfrak{h})$ is freely generated by $\mathfrak{h}^*[-1]$ as a polynomial algebra, a morphism on the left amounts to a graded linear map
$ \mathfrak{h}^*[-1]\to \mathrm{CE}(\hat{\mathfrak{g}})$ constrained by the compatibility with the differentials condition. As $\mathrm{CE}(\hat{\mathfrak{g}})=\mathrm{CE}(\mathfrak{g})[y_1]$,
where $y_1$ is a variable in degree 1 with $d_{\hat{\mathfrak{g}}}y_1=f^*(x_2)$, as a graded vector space we have
\[
\mathrm{CE}(\hat{\mathfrak{g}})=\mathrm{CE}(\mathfrak{g})\oplus y_1\, \mathrm{CE}(\mathfrak{g})=\mathrm{CE}(\mathfrak{g})\oplus \mathrm{CE}(\mathfrak{g})[-1],
\]
so that a graded linear map $ \mathfrak{h}^*[-1]\to \mathrm{CE}(\hat{\mathfrak{g}})$ is equivalent to a pair of graded linear maps from $ \mathfrak{h}^*[-1]$ to $\mathrm{CE}(\mathfrak{g})$ and to $\mathrm{CE}(\mathfrak{g})[-1]$, respectively. In turn, this pair is a graded linear map $\mathfrak{h}^*[-1]\oplus  \mathfrak{h}^*\to \mathrm{CE}(\mathfrak{g})$. We can extend this to a graded linear map
\[
\mathfrak{h}^*[-1]\oplus  \mathfrak{h}^*\oplus b\mathfrak{u}_1^*[-1]
\longrightarrow \mathrm{CE}(\mathfrak{g})
\]
by mapping the linear generator $x_2$ of $b\mathfrak{u}_1^*[-1]$ to the element $f^*(x_2)$ of $\mathrm{CE}(\mathfrak{g})$. This way we define a graded commutative algebra map
\[
\Exterior^\bullet(\mathfrak{h}^*[-1]\oplus  \mathfrak{h}^*\oplus b\mathfrak{u}_1^*[-1])
\longrightarrow \mathrm{CE}(\mathfrak{g})
\]
which a direct computation shows to be a morphism of DGCAs making the diagram 
\[
\xymatrix@R=1.3em{
& \mathbb{R}[x_2]\ar[rd]^{f^*}\ar[ld]\\
\mathrm{CE}(\mathrm{cyc}(\mathfrak{h}))\ar[rr]&&\mathrm{CE}(\mathfrak{g})
}
\]
commute. See \cite{FSS16} for details.
\begin{example}
The Sullivan model for $S^4$ is 
\[
\mathrm{CE}(\mathfrak{l}S^4)=(\mathbb{R}[z_4,z_7],\ dz_4=0,\, dz_7={z_4}^2).
\]
Therefore, the Sullivan model for $\mathcal{L}S^4/\!/U(1)$ is
\[
\mathrm{CE}(\mathrm{cyc}(\mathfrak{l}S^4))=(\mathbb{R}[x_2,y_3,z_4,y_6,z_7],\ dx_2=0,\, dy_3=0,\, dz_4=y_3x_2,\, dy_6=-2y_3z_4,\, dz_7={x_4}^2+x_2y_6).
\]
Making the change of variables $f_2=x_2$, $h_3=y_3$, $f_4=z_4$, $f_6=-\frac{1}{2}y_6$, and $h_7=z_7$, this can be rewritten as
\[
\mathrm{CE}(\mathrm{cyc}(\mathfrak{l}S^4))=(\mathbb{R}[f_2,f_4,f_6,h_3,h_7],\ df_2=0,\, dh_3=0,\, df_4=h_3f_2,\, df_6=y_3t_4,\, dh_7={f_4}^2-2t_2t_6).
\]
Therefore, a smooth cocycle $X\to \mathrm{cyc}(\mathfrak{l}S^4)$ on a smooth (super)manifold $X$ will be the datum of a closed 3-form $H_3$ and of 2-, 4- and 6-forms $F_2$, $F_4$ and $F_6$ on $X$ such that
\[
dF_2=0;\qquad dF_4=H_3\wedge F_2;\qquad dF_6= H_3\wedge F_4,
\]
together with a 7-form $H_7$ which is a potential for the closed 8-form $F_4\wedge F_4-2F_2\wedge F_6$. In particular, if $Y\to X$ is rationally a principal $S^1$-bundle, then a $\mathfrak{l}S^4$ cocycle on $Y$ will induce, by the hofiber/cyclification adjunction, such a set of differential forms on $X$. Notice in particular how the above equations for the differentials of the $F_{2n}$'s are precisely (a subset of) the equations for a $H_3$-twisted cocycle $\sum_{n=-\infty}^\infty F_{2n}u^n$ in $(\Omega^\bullet(X)[[u^{-1},u]],d_{H_3})$ with $F_0=0$. This is the mechanism by which the M-theory cocycle $\R^{10, 1|{\bf 32}}\to \mathfrak{l}S^4$ induces  twisted (rational) even K-theory cocycles on on $\mathbb{R}^{9,1\vert \mathbf{16}+ \overline{\mathbf{16}}}$; see  \cite{FSS16a}.
\end{example}

\paragraph{Fiber integration revisited.}
The $L_\infty$ algebras $b^n\mathfrak{u}_1$ have a particularly simple cyclification. Namely, as $\mathrm{CE}(b^n\mathfrak{u}_1)=(\mathbb{R}[x_{n+1}],0)$, we see from the explicit description of cyclification given in the previous section that as a polynomial algebra
$\mathrm{CE}(\mathrm{cyc}(b^n\mathfrak{u}_1))$ is obtained from $\mathbb{R}[x_{n+1}]$ by adding a generator $y_n=sx_{n+1}$ in degree $n$ and a generator $z_2$ in degree 2. The differential is given by
\[
dx_{n+1}=z_2\, y_n; \qquad dy_n=0; \qquad dz_2=0.
\]
From this one immediately sees that we have an injection $(\mathbb{R}[y_n],0)\hookrightarrow (\mathrm{CE}(\mathrm{cyc}(b^n\mathfrak{u}_1)),d)$ and so dually a fibration
\[
\mathrm{cyc}(b^n\mathfrak{u}_1)\longrightarrow b^{n-1}\mathfrak{u}_1
\]
of $L_\infty$-algebras. Then given an $\mathbb{R}$ central extension $\pi\colon \hat{\mathfrak{g}}\to\mathfrak{g}$ we can form the composition of morphisms of sets
\[
\mathrm{Hom}_{L_\infty}(\hat{\mathfrak{g}},b^n\mathfrak{u}_1)\cong \mathrm{Hom}_{L_\infty/b\mathfrak{u}_1}(\mathfrak{g},\mathrm{cyc}(b^n\mathfrak{u}_1))\to
\mathrm{Hom}_{L_\infty}(\mathfrak{g},\mathrm{cyc}(b^n\mathfrak{u}_1))\to \mathrm{Hom}_{L_\infty}(\mathfrak{g},b^{n-1}\mathfrak{u}_1),
\]
and a direct inspection easily reveals that this coincides with the fiber integration morphism 
\[
\pi_*\colon \mathrm{Hom}_{L_\infty}(\hat{\mathfrak{g}},b^n\mathfrak{u}_1)
\longrightarrow
\mathrm{Hom}_{L_\infty}(\mathfrak{g},b^{n-1}\mathfrak{u}_1)
\]
from Section \ref{fiber-integration}.

\section{Rational homotopy theory of T-duality configurations}
\label{rht-tfold}

\subsection{The classifying spaces of T-duality configurations}
As we already noticed, the same way as the classifying space $BU(1)$ of principal $U(1)$-bundles is a $K(\mathbb{Z},2)$, the classifying space $B^3U(1)$ of principal $U(1)$-2-bundles (or principal $U(1)$-2-gerbes) is a $K(\mathbb{Z};4)$. This implies that the cup product map
\[
\cup \colon K(\mathbb{Z},2)\times K(\mathbb{Z},2)\longrightarrow K(\mathbb{Z},4)
\] 
is equivalently a map
\[
\cup \colon BU(1)\times BU(1)\longrightarrow B^3U(1),
\] 
i.e., to any pair of principal $U(1)$ bundles $P_1$ and $P_2$ on a manifold $X$ is canonically associated a $U(1)$-$2$-gerbe $P_1\cup P_2$ on $X$. By definition, a topological T-duality configuration is the datum of two such principal $U(1)$-bundles together with a trivialization of their cup product. In other words, a topological T-duality configuration on a manifold $X$ is a homotopy commutative diagram
\[
\xymatrix@=1.5em{
X\ar[rr]\ar[d]&&{*}\ar[d]\\
BU(1)\times BU(1)\ar[rr]^-{\cup}&&B^3U(1)
\;.}
\]
By the universal property of the homotopy pullback this is in turn equivalent to a map from $X$ to the homotopy fiber of the cup product, which will therefore be the classifying space for topological T-duality configurations. To fix notations, let us call $BTfolds$ this classifying space, so that $BTfold$ is defined by the homotopy pullback
\[
\xymatrix@=1.5em{
B{\rm Tfold}\ar[rr]\ar[d]&&{*}\ar[d]\\
BU(1)\times BU(1)\ar[rr]^-{\cup}&&B^3U(1)
\;.}
\]
The rationalization of $B{\rm Tfold}$ is obtained as the $L_\infty$-algebra $b\mathfrak{tfold}$ given by the homotopy pullback
\[
\xymatrix@R=1.5em{
b\mathfrak{tfold}\ar[rr]\ar[d]&&{0}\ar[d]\\
b\mathfrak{u}_1\times b\mathfrak{u}_1\ar[rr]^-{\cup}&&b^3\mathfrak{u}_1
\;,
}
\]
and in order to get an explicit description of it we only need to give an explicit description of the 4-cocycle $b\mathfrak{u}_1\times b\mathfrak{u}_1\xrightarrow{\cup}b^3\mathfrak{u}_1$. This is easily read in the dual picture: it is the obvious morphism of CGDAs 
\begin{align*}
(\mathbb{R}[x_4],0)&\longrightarrow (\mathbb{R}[\check{x}_2,\tilde{x}_2],0)\cong (\mathbb{R}[x_2],0)\otimes (\mathbb{R}[x_2],0) \\
x_4&\longmapsto \check{x}_2\, \tilde{x}_2.
\end{align*}
The Chevalley-Eilenberg algebra of $b\mathfrak{tfold}$ is then given by the homotopy pushout
\[
\xymatrix@=1.6em{
(\mathbb{R}[x_4],0)\ar[rr]\ar[d]_-{\cup^*}&&({\mathbb{R}},0)\ar[d]\\
(\mathbb{R}[\check{x}_2,\tilde{x}_2],0)\ar[rr]&&(\mathrm{CE}(b\mathfrak{tfold}),d)
\;,
}
\]
i.e., by the pushout
\[
\xymatrix@=1.6em{
(\mathbb{R}[x_4],0)\ar[rr]\ar[d]_-{\cup^*}&&({\mathbb{R}}[y_3,x_4],dy_3=x_4)\ar[d]\\
(\mathbb{R}[\check{x}_2,\tilde{x}_2],0)\ar[rr]&&(\mathrm{CE}(b\mathfrak{tfold}),d)
\;.
}
\]
Explicitly, this means that
\[
(\mathrm{CE}(b\mathfrak{tfold}),d)=(\mathbb{R}[\check{x}_2,\tilde{x}_2,y_3], d\check{x}_2=0, d\tilde{x}_2=0, dy_3= \check{x}_2\, \tilde{x}_2), 
\]
and so an $L_\infty$-morphism $\mathfrak{g}\to b\mathfrak{tfold}$ is precisely what we should have expected it to be: a pair of 2-cocycles on $\mathfrak{g}$ together with a trivialization of their product. Moreover, one manifestly has an isomorphism
\[
(\mathrm{CE}(b\mathfrak{tfold}),d)\cong (\mathrm{CE}(\mathrm{cyc}(b^2\mathfrak{u}_1),d)
\]
so that the $b\mathfrak{tfold}$ $L_\infty$-algebra is isomorphic to the cyclification of $b^2\mathfrak{u}_1$. This result actually already holds at the topological level, i.e., there is a homotopy equivalence $BTfold\cong \mathrm{cyc}(K(\mathbb{Z},3))
\cong \mathrm{cyc}(B^2U(1))$. Proving this equivalence beyond the rational approximation is however harder; see \cite{BunkeSchick05} for a proof.

\medskip
The $L_\infty$-algebra $b\mathfrak{tfold}$ has two independent 2-cocycles $f_1,f_2\colon b\mathfrak{tfold}\to b\mathfrak{u}_1$ given in the dual picture by $f_1^*(x_2)=\check{x}_2$ and by $f_2^*(x_2)=\tilde{x}_2$. Let us denote by $\mathfrak{p}_1$ and $\mathfrak{p_2}$ the central extensions of $b\mathfrak{tfold}$ corresponding to $f_1$ and $f_2$, respectively. They are clearly isomorphic as $L_\infty$-algebras; however they are not equivalent as $L_\infty$-algebras over $b\mathfrak{tfold}$ as the two classifying morphisms $f_1$ and $f_2$ are not homotopy equivalent.

\medskip
Let us now write $\mathbb{R}[x_3]$ for the Chevalley-Eilenberg algebra $\mathrm{CE}(b^2\mathfrak{u}_1)$, so that in the notation of Section \ref{cyc-l-infty} we have $\mathrm{CE}(\mathrm{cyc}(b^2\mathfrak{u}_1))=\mathbb{R}[x_3,y_2,z_2]$ with $dx_3= z_2y_2$, $dy_2=0$ and $dz_2=0$, and with the canonical 2-cocycle $\mathrm{cyc}(b^2\mathfrak{u}_1)\to b\mathfrak{u}_1$ being given dually by
\begin{align*}
f_{\mathrm{cyc}}^*\colon \mathbb{R}[x_2]&\longrightarrow  \mathbb{R}[x_3,y_2,z_2]\\
x_2&\longmapsto z_2.
\end{align*}
The isomorphism of $L_\infty$-algebras $\varphi_1\colon b\mathfrak{tfold}\to \mathrm{cyc}(b^2\mathfrak{u}_1)$ dually given by $x_3\mapsto y_3$, $y_2\mapsto \tilde{x}_2$ and $z_2\mapsto \check{x}_2$ is such that the diagram of DGCAs
\[
\xymatrix@=1.5em{
& \mathrm{CE}(b\mathfrak{u}_1)\ar[rd]^{f_1^*}\ar[dl]_{f^*_{\mathrm{cyc}}}\\
\mathrm{CE}(\mathrm{cyc}(b^2\mathfrak{u}_1))\ar[rr]^{\varphi_1^*}&&\mathrm{CE}(b^2\mathfrak{tfold})
}
\]
commutes, i.e., $\varphi_1$ is an isomorphism over $b\mathfrak{u}_1$. Hence, by the hofiber/cyclification adjunction, it corresponds to an $L_\infty$ morphism from the homotopy fiber of $f_1$ to $b^2\mathfrak{u}_1$, i.e., to a 3-cocycle $a_{3,1}$ over $\mathfrak{p}_1$. Repeating the same reasoning for $f_2$ we get a canonical 3-cocycle $a_{3,2}$ over $\mathfrak{p_2}$. Therefore, we see how some of the ingredients of a rational T-duality configuration naturally emerge form the T-fold $L_\infty$-algebra. The cocycles $a_{3,1}$ and $a_{3,2}$ can be easily given an explicit description, by unwinding the hofiber/cyclification adjunction in this case. Let us do this for $a_1$. The homotopy fiber $\mathfrak{p}_1$ of $f_1$ is defined by the homotopy of DGCAs
\[
\xymatrix{
(\mathbb{R}[x_2],0)\ar[rr]\ar[d]_{f_1^*}&&(\mathbb{R},0)\ar[d]\\
(\mathbb{R}[\check{x}_2,\tilde{x}_2,y_3],d\check{x}_2=d\tilde{x}_2=0, dy_3=\check{x}_2\tilde{x}_2)\ar[rr]&&(\mathrm{CE}(\mathfrak{p}_1),d_{\mathfrak{p}_1})
\;,
}
\]
and so it is given by
\[
(\mathrm{CE}(\mathfrak{p}_1),d_{\mathfrak{p}_1})=(\mathbb{R}[\check{y}_1,\check{x}_2,\tilde{x}_2,y_3], d\check{y}_1=\check{x}_2, d\check{x}_2=d\tilde{x}_2=0,dy_3=\check{x}_2\tilde{x}_2 ).
\]
One immediately sees the relation
\[
dy_3=d(\check{y}_1\tilde{x_2}),
\]
i.e., that $y_3-\check{y}_1\tilde{x_2}$ is a 3-cocycle on $\mathfrak{p}_1$. Under the hofiber/cyclification adjunction this 3-cocycle corresponds to the morphism of DGCAs $\mathrm{CE}(\mathrm{cyc}(b^2\mathfrak{u}_1))\to \mathrm{CE}(b^2\mathfrak{tfold})$ mapping $x_3$ to $y_3$, $y_2$ to $\tilde{x}_2$ and $z_2$ to $\check{x}_2$, i.e., to the morphism $\varphi_1$. In other words, 
\[
a_{3,1}=y_3-\check{y}_1\tilde{x_2}.
\]
In a perfectly similar way $a_{3,2}=y_3-\check{x}_2\tilde{y}_1$. Finally, let us form the homotopy fiber product $\mathfrak{t}=\mathfrak{p}_1\times_{b\mathfrak{tfold}}\mathfrak{p}_2$. It is described by the Chevalley-Eilenberg algebra
\[
(\mathrm{CE}(\mathfrak{t}),d_{\mathfrak{t}})=(\mathbb{R}[\check{y}_1,\tilde{y}_1,\check{x}_2,\tilde{x}_2,y_3], d\check{y}_1=\check{x}_2, d\tilde{y}_1=\tilde{x}_2,dy_3=\check{x}_2\tilde{x}_2), 
\]
with the projections $\pi_i\colon \mathfrak{t}\to \mathfrak{p}_i$ given in the dual picture by the obvious inclusions. By construction, $\pi_1$ and $\pi_2$ are $\mathbb{R}$-central extensions, classified by the 2-cocycles $\tilde{x}_2$ and $\hat{x}_2$, respectively. One computes
\[
\pi_1^*a_{3,1}-\pi_2^*a_{3,2}=(y_3-\check{y}_1\tilde{x_2})-(y_3-\check{x}_2\tilde{y}_1)=-\check{y}_1\tilde{x_2}+\check{x}_2\tilde{y}_1=-\check{y}_1(d\tilde{y}_1)+(d\check{y}_1)\tilde{y_1}=d(\check{y}_1\tilde{y}_1),
\]
i.e.,
\[
\pi_1^*a_{3,1}-\pi_{3,2}^*a_2=db_2,
\]
where $b_2\in \mathrm{CE}(\mathfrak{t})$ is the degree 2 element $b=\check{y}_1\tilde{y}_1$. Thus we see that the $L_\infty$-algebra $b\mathfrak{tfold}$ actually contains all the data of a quintuple $(\pi_1,\pi_2,a_{3,1},a_{3,2},b_2)$ inducing a Fourier-Mukai transform.

\subsection{Maps to $b\mathfrak{tfold}$}\label{rational-tfold-config}
 All of the construction of the quintuple $(\pi_1,\pi_2,a_1,a_2,b)$ out of the the $L_\infty$-algebra $b\mathfrak{tfold}$ can be pulled back along a morphism of $L_\infty$-algebras $\mathfrak{g}\to b\mathfrak{tfold}$. That is, given such a morphism one has two $\mathbb{R}$-central extensions $\mathfrak{g}_1$ and $\mathfrak{g}_2$ of $\mathfrak{g}$ together with 3-cocycles $a_{3,1}$ and $a_{3,2}$ on $\mathfrak{g}_1$ and $\mathfrak{g}_2$, respectively, and a degree 2 element $b_2$ on the (homotopy) fiber product $L_\infty$-algebra $\mathfrak{g}_1\times_{\mathfrak{g}}\mathfrak{g}_2$ with $\pi_1^*a_{3,1}-\pi_2^*a_{3,2}=db_2$. Let us see in detail how this works. 
 
 \medskip
 To begin with, the datum of a morphism $\mathfrak{g}\to b\mathfrak{tfold}$ is precisely the datum of two 2-cocycles $\check{c}_2$ and $\tilde{c}_2$ on $\mathfrak{g}$ together with a degree 3 element $h_3\in \mathrm{CE}(\mathfrak{g})$ such that $dh_3=\check{c}_2\tilde{c}_2$. The two cocycles $\check{c}_2$ and $\tilde{c}_2$  define the two $\mathbb{R}$-central extensions  $\mathfrak{g}_1$ and $\mathfrak{g}_2$ of  $\mathfrak{g}$ defined, respectively, by
\begin{align*}
(\mathrm{CE}(\mathfrak{g}_1),d_{\mathfrak{g}_1})&=(\mathrm{CE}(\mathfrak{g})[\check{e}_1], d \check{e}_1=\check{c}_2)\;,
\\
(\mathrm{CE}(\mathfrak{g}_2),d_{\mathfrak{g}_2})&=(\mathrm{CE}(\mathfrak{g})[\tilde{e}_1], d \tilde{e}_1=\tilde{c}_2)\;.
\end{align*}
On the $L_\infty$-algebra $\mathfrak{g}_1$ we have the 3-cocycle $a_{3,1}=h_3-\check{e}_1\tilde{c}_2$, and on the $L_\infty$-algebra $\mathfrak{g}_2$ we have the 3-cocycle $a_{3,2}=h_3-\check{c}_2\tilde{e}_1$. Finally, the homotopy fiber product $\mathfrak{g}_1\times_{\mathfrak{g}}\mathfrak{g}_2$ is given by
\[
\big(\mathrm{CE}(\mathfrak{g}_1\times_{\mathfrak{g}}\mathfrak{g}_2),d_{\mathfrak{g}_1\times_{\mathfrak{g}}\mathfrak{g}_2}\big)
=\big(\mathrm{CE}(\mathfrak{g})[\check{e}_1,\tilde{e}_1]; \; 
d \check{e}_1=\check{c}_2, d \tilde{e}_1=\tilde{c}_2\big)\;,
\]
and so in $\mathrm{CE}(\mathfrak{g}_1\times_{\mathfrak{g}}\mathfrak{g}_2)$ we have $\pi_1^*a_{3,1}-\pi_2^*a_{3,2}=db_2$, where $\pi_1^*$ and $\pi_2^*$ are the obvious inclusions and $b_2=\check{e}_1\tilde{e}_1$. Notice that $\mathrm{CE}(\mathfrak{g}_1\times_{\mathfrak{g}}\mathfrak{g}_2)$ is built from $\mathrm{CE}(\mathfrak{g}_1)$ by adding the additional generator $\tilde{e}_1$ and from $\mathrm{CE}(\mathfrak{g}_2)$ by adding the additional generator $\check{e}_1$. We can now make completely explicit the Fourier-Mukai transform
\[
\Phi_{b_2}\colon H_{L_\infty;a_{3,1}}^{{\bullet}}(\mathfrak{g}_1)[[u^{-1},u]] 
\longrightarrow
H_{L_\infty;a_{3,2}}^{{\bullet}-{1}}(\mathfrak{g}_2)[[u^{-1},u]].
\]
To fix notation, let
\[
\xymatrix@=1.5em{
&&\mathfrak{g}_1\times_{\mathfrak{g}}\mathfrak{g}_2
\ar[lld]_-{\pi_1}\ar[rrd]^-{\pi_2}\\
\mathfrak{g}_1\ar[drr]_-{p_1}&&&&\mathfrak{g}_2\ar[dll]^-{p_2}\\
&&\mathfrak{g}
}
\]
be the homotopy fiber product defining $\mathfrak{g}_1\times_{\mathfrak{g}}\mathfrak{g}_2$. Notice that the Beck-Chevalley condition
\footnote{See \cite{MV} for a general discussion of this condition for proper maps of toposes.}
 \begin{equation}
 \label{eq.beck-chevalley}p_2^*p_{1*}=\pi_{2*}\pi_1^*
 \end{equation}
 holds.
Indeed, for any $\omega_k=\alpha_k+\check{e}_1\beta_{k-1}$ in $\mathrm{CE}(\mathfrak{g}_1)$, we have
\[
\pi_{2*}\pi_1^*\omega_k=\pi_{2*}\pi_1^*(\alpha_k+\check{e}_1\beta_{k-1})=\pi_{2*}(\alpha_k+\check{e}_1\beta_{k-1})=\beta_{k-1}=p_{1*}\omega_k=p_2^*p_{1*}\omega_k\;.
\]
Let us write $\omega_{2n}=\alpha_{2n}+\check{e}_1\beta_{2n-1}$ for a degree $2n$ element in $\mathrm{CE}(\mathfrak{g}_1)$ and $
\omega=\sum_{n\in \mathbb{Z}}u^{k-n}\omega_{2n}
$ for a degree $2k$ element in $\omega\in \mathrm{CE}(\mathfrak{g}_1)[[u^{-1},u]]$.  The Fourier-Mukai transform $\Phi_{b_2}$ maps the element $\omega$ to $\pi_{2*}(e^{b_2}\pi_{1}^*\omega)$. Since $\pi_1^*$ is just the inclusion and 
$
e^{u^{-1}b_2}=e^{u^{-1}\check{e}_1\tilde{e}_1}= 1 + u^{-1}\check{e}_1\tilde{e}_1
$, we find
\begin{align*}
\Phi_{b_2}(\omega)&=\pi_{2*}(\omega+ u^{-1}\check{e}_1\tilde{e}_1\omega)\\
&=\sum_{n\in \mathbb{Z}}u^{k-n}\pi_{2*}(\alpha_{2n}+\check{e}_1\beta_{2n-1}+ u^{-1}\check{e}_1\tilde{e}_1(\alpha_{2n}+\check{e}_1\beta_{2n-1}))\\
&=\sum_{n\in \mathbb{Z}}u^{k-n}\pi_{2*}(\alpha_{2n}+\check{e}_1\beta_{2n-1}+ u^{-1}\check{e}_1\tilde{e}_1\alpha_{2n})\\
&=\sum_{n\in \mathbb{Z}}u^{k-n}(\beta_{2n-1}+\tilde{e}_1\alpha_{2n-2})\;.
\end{align*}
Let $\tilde{\omega}_{2n-1}=\beta_{2n-1}+\tilde{e}_1\alpha_{2n-2}$ and $\tilde{\omega}=\sum_{n\in \mathbb{Z}}u^{k-n}\tilde{\omega}_{2n-1}$, so that $\tilde{\omega}$ is a degree  $2k-1$ element in $\mathrm{CE}(\mathfrak{g}_2)[[u^{-1},u]]$ and $ \tilde{\omega}=\Phi_{b_2}(\omega)$.  We know from the general construction of Fourier-Mukai transforms we have been developing that if $\omega$ is an an $a_{3,1}$-twisted cocycle, then $\tilde{\omega}$ is an $a_{3,2}$-twisted cocycle. We can directly show this as follows.
The degree $2k$ cochain $\omega$ is a $a_{3,1}$-twisted degree $2k$ cocycle precisely when
\[
d_{\mathfrak{g}_1}\omega+u^{-1}a_{3,1}\,\omega=0.
\]
This equation is in turn equivalent to the system of equations
\[
d_{\mathfrak{g}_1}\omega_{2n}+ a_{3,1}\omega_{2n-2}=0, \qquad n\in \mathbb{Z}.
\]
Writing 
$
\omega_{2n}=\alpha_{2n}+\check{e}_1\beta_{2n-1}
$
and recalling that 
$
a_{3,1}=h_3-\check{e}_1\tilde{c}_2
$,
this becomes
\[
d_{\mathfrak{g}}\alpha_{2n}+\check{c}_2\beta_{2n-1}-\check{e}_1d_{\mathfrak{g}}\beta_{2n-1}+h_3\alpha_{2n-2}-\check{e}_1\tilde{c}_2\alpha_{2n-2}-\check{e}_1h_3\beta_{2n-3}=0,
\]
i.e.,
\[
\begin{cases}
d_{\mathfrak{g}}\alpha_{2n}+h_3\alpha_{2n-2}=\check{c}_2\beta_{2n-1},\\
d_{\mathfrak{g}}\beta_{2n-1}+h_3\beta_{2n-3}=\tilde{c}_2\alpha_{2n-2}.
\end{cases}
\]
Then we can compute
\begin{align*}
d_{\mathfrak{g}_2}\tilde{\omega}_{2n-1}&=d_{\mathfrak{g}_2}(\beta_{2n-1}+\tilde{e}_1\alpha_{2n-2})\\
&=d_{\mathfrak{g}}\beta_{2n-1}+\tilde{c}_2\alpha_{2n-2}-\tilde{e}_1d_{\mathfrak{g}}\alpha_{2n-2}\\
&=(-h_3\beta_{2n-3}-\tilde{c}_2\alpha_{2n-2})+\tilde{c}_2\alpha_{2n-2}-\tilde{e}_1(-h_3\alpha_{2n-4}-\check{c}_2\beta_{2n-3})\\
&=-a_{3,2}\beta_{2n-3}+\tilde{e}_1a_{3;2}\alpha_{2n-4}\\
&=-a_{3,2}(\beta_{2n-3}+\tilde{e}_1\alpha_{2n-4})\\
&=-a_{3,2}\tilde{\omega}_{2n-3}\;,
\end{align*}
which shows that $\tilde{\omega}$ is a degree $2k-1$ $a_{3,2}$-twisted cocycle.

\medskip
Looking at the explicit formula for $\Phi_{b_2}$ we have now determined above, we see that $\Phi_{b_2}$ acts as
\[
\sum_{n\in \mathbb{Z}}u^{k-n}(\alpha_{2n}+\check{e}_1\beta_{2n-1})\longmapsto\sum_{n\in \mathbb{Z}}u^{k-n}(\beta_{2n-1}+\tilde{e}_1\alpha_{2n-2})\;.
\]
So it is manifestly a linear isomorphism between the space of degree $2k$ cochains in $\mathrm{CE}(\mathfrak{g}_1)[[u^{-1},u]]$ and degree $2k-1$ cochains in $\mathrm{CE}(\mathfrak{g}_2)[[u^{-1},u]]$. Repeating verbatim the above argument one sees that $\Phi_{b_2}$ is also a linear isomorphism between degree $2k-1$ cochains in $\mathrm{CE}(\mathfrak{g}_1)[[u^{-1},u]]$ and degree $2k-2$ cochains in $\mathrm{CE}(\mathfrak{g}_2)[[u^{-1},u]]$.
 Not surprisingly, the inverse morphism is
$u\Phi_{-b_2}$  in both cases. This can be showed directly by repeating once more the argument above, or specializing to a rational T-duality configuration the general formula for the composition of two Fourier-Mukai transforms. We are going to show this in the following section. Either way, as $\Phi_{b_2}$ is also a morphism of complexes, it is an isomorphism of complexes and so, in particular one sees that the Fourier-Mukai transform associated to an $L_\infty$-morphism $\mathfrak{g}\to b\mathfrak{tfold}$ is an isomorphism
\[
\Phi_{b_2}\colon H_{L_\infty;a_{3,1}}^{\bullet}(\mathfrak{g}_1;\mathbb{R})[[u^{-1},u]] 
\xrightarrow{\;\sim\;} H_{L_\infty;a_{3,2}}^{\bullet-1}(\mathfrak{g}_2;\mathbb{R})[[u^{-1},u]]\;.
\]

\subsection{Compositions of Fourier-Mukai transforms}
Finally, let us describe the composition of Fourier-Mukai transforms. To that end, we will consider 
a pair of quintuple $(\pi_1,\pi_2,a_{3,1},a_{3,2},b_2)$ and $(\tilde{\pi}_1,\tilde{\pi}_2,{a}_{3,2},{a}_{3,3},\tilde{b}_2)$, which induce two corresponding Fourier-Mukai transforms $\Phi_{b_2}\colon H_{L_\infty;a_{3,1}}^{\bullet}(\mathfrak{g}_1;\mathbb{R})[[u^{-1},u]] \to H_{L_\infty;a_{3,2}}^{\bullet-1}(\mathfrak{g}_2;\mathbb{R})[[u^{-1},u]]$ and $\Phi_{\tilde{b}_2}\colon H_{L_\infty;a_{3,2}}^{\bullet}(\mathfrak{g}_2;\mathbb{R})[[u^{-1},u]] \to H_{L_\infty;a_{3,3}}^{\bullet-1}(\mathfrak{g}_3;\mathbb{R})[[u^{-1},u]]$, respectively. To describe the composition $\Phi_{\tilde{b}_2}\circ\Phi_{b_2}$, we form the fiber product $\mathfrak{h}_1\times_{\mathfrak{g}_2}\mathfrak{h}_2$, where $\mathfrak{h}_1$ and $\mathfrak{h}_2$ are the $L_\infty$ algebras appearing as ``roofs'' in the spans defining $\Phi_{\tilde{b}_2}$ and $\Phi_{b_2}$, respectively. Notice that, as $\pi_2\colon \mathfrak{h_1}\to\mathfrak{g}_2$ and $\tilde{\pi}_1\colon \mathfrak{h_2}\to\mathfrak{g}_2$ are fibrations, $\mathfrak{h}_1\times_{\mathfrak{g}_2}\mathfrak{h}_2$ is actually a model for the homotopy fiber product of $\mathfrak{h}_1$ and $\mathfrak{h}_2$ over $\mathfrak{g}_2$. Then we have the diagram
\[
\xymatrix{&&\mathfrak{h}_1\times_{\mathfrak{g}_2}\mathfrak{h}_2\ar[ld]_{q_1}\ar[rd]^{q_2}\ar@/^2.5pc/[rrdd]^{p_2}\ar@/_2.5pc/[lldd]_{p_1}\\
& \mathfrak{h}_1\ar[dl]_{\pi_1}\ar[dr]^{\pi_2}&&\mathfrak{h}_2\ar[ld]_{\tilde{\pi}_1}\ar[rd]^{\tilde{\pi}_2}\\
\mathfrak{g}_1&&\mathfrak{g}_2&& \mathfrak{g}_3
\;,
}
\]
where $q_1$ and $q_2$ are the projections, and where $p_1=\pi_1 q_1$ and 
$p_2=\tilde{\pi}_2q_2$. By definition of Fourier-Mukai transform and by the 
Beck-Chevalley condition (equation \eqref{eq.beck-chevalley})
$\tilde{\pi}_1^*\pi_{2*}=q_{2*}q_1^*$, for any $\omega$ in $\mathrm{CE}(\mathfrak{g}_1)$ we have
\begin{align*}
(\Phi_{\tilde{b}_2}\circ\Phi_{b_2})(\omega)&=\tilde{\pi}_{2*}(e^{u^{-1}\tilde{b}_2}\tilde{\pi}_1^*\pi_{2*}(e^{u^{-1}b_2}\pi^*\omega))\\
&=\tilde{\pi}_{2*}(e^{u^{-1}\tilde{b}_2}q_{2*}q_1^*(e^{u^{-1}b_2}\pi_1^*\omega))\\
&=\tilde{\pi}_{2*}(e^{u^{-1}\tilde{b}_2}q_{2*}q_1^*(e^{u^{-1}b_2}\pi_1^*\omega))\\
&=\tilde{\pi}_{2*}(e^{u^{-1}\tilde{b}_2}q_{2*}(q_1^*e^{u^{-1}b_2}\, q_1^*\pi^*\omega))\\
&=\tilde{\pi}_{2*}(e^{u^{-1}\tilde{b}_2}q_{2*}(e^{u^{-1}q_1^*b_2}\, p_1^*\omega)).
\end{align*}
Now recall the projection formula (equation \ref{eq.projection}), and use the fact that  to get $e^{u^{-1}\tilde{b}_2}$ entirely consists of even components to get
\[
q_{2*}(q_2^*(e^{u^{-1}\tilde{b}_2})\,  e^{u^{-1}q_1^*b_2}\, p_1^*\omega)=e^{u^{-1}\tilde{b}_2}\, q_{2*}(e^{u^{-1}q_1^*b_2}\, p_1^*\omega).
\]
Therefore,
\begin{align*}
(\Phi_{\tilde{b}_2}\circ\Phi_{b_2})(\omega)&=\tilde{\pi}_{2*}q_{2*}(q_2^*(e^{u^{-1}\tilde{b}_2})\,  e^{u^{-1}q_1^*b_2}\, p_1^*\omega)\\
&=p_{2*}(e^{u^{-1}(q_2^*\tilde{b}_2+q_1^*b_2)}\, p_1^*\omega).
\end{align*}
By definition of fiber product, the two morphisms $q_2^*\tilde{\pi}_1^*$ and $q_1^*\pi_2^*$ coincide. Therefore,
\begin{align*}
d_{\mathfrak{h}_1\times_{\mathfrak{g}_2}\mathfrak{h}_2}(q_2^*\tilde{b}_2+q_1^*b_2)&=q_2^*d_{\mathfrak{h}_2}\tilde{b}_2+q_1^*d_{\mathfrak{h}_1}b_2\\
&=q_2^*(\tilde{\pi}_1^*a_{3,2}-\tilde{\pi}_2^*a_{3,3})+q_1^*(\pi_1^*a_{3,1}-\pi_2^*a_{3,2})\\
&=q_1^*(\pi_1^*a_{3,1})-q_2^*(\tilde{\pi}_2^*a_{3,3})+(q_2^*\tilde{\pi}_1^*-q_1^*\pi_2^*)a_{3,2}\\
&=p_1^*a_{3,1}-p_2^*a_{3,3}.
\end{align*}
This shows that $\Phi_{\tilde{b}_2}\circ\Phi_{b_2}$ is indeed the Fourier-Mukai transform associated with the quintuple $(p_1,p_2,a_{3,1},a_{3,3},q_1^*b_2+q_2^*\tilde{b}_2)$. We write this as
\[
\Phi_{\tilde{b}_2}\circ\Phi_{b_2}=\Phi_{q_1^*b_2+q_2^*\tilde{b}_2}.
\]
Notice that $p_1\colon \mathfrak{h}_1\times_{\mathfrak{g}_2}\mathfrak{h}_2\to\mathfrak{g}_1$ and $p_2\colon \mathfrak{h}_1\times_{\mathfrak{g}_2}\mathfrak{h}_2\mathfrak{g}_3$ are not $\mathfrak{u}_1$-central extensions but $\mathfrak{u}_1\times \mathfrak{u}_1$-central extensions, so the Fourier-Mukai transform 
$\Phi_{q_1^*b_2+q_2^*\tilde{b}_2}$ lowers the degree by 2.

\medskip
It is interesting to specialize this to the case where $(\pi_1,\pi_2,a_{3,1},a_{3,2},b_2)$ is the 
quintuple associated with a  rational T-duality configuration $\mathfrak{g}\to b\mathfrak{tfold}$ and $(\tilde{\pi}_1,\tilde{\pi}_2,{a}_{3,2},{a}_{3,3},\tilde{b}_2)=(\pi_2,\pi_1,{a}_{3,2},{a}_{3,1},-b_2)$. In this case
\[
\big(\mathrm{CE}(\mathfrak{h}_1\times_{\mathfrak{g}_2}\mathfrak{h}_2),d_{\mathfrak{h}_1\times_{\mathfrak{g}_2}\mathfrak{h}_2}\big)
=\big(\mathrm{CE}(\mathfrak{g})[\check{e}_{1,1},\tilde{e}_1,\check{e}_{1,2}]; \; d \check{e}_{1,1}=d \check{e}_{1,2}=\check{c}_2, d \tilde{e}_1=\tilde{c}_2\big)\;,
\]
and the morphisms $q_i^*\colon \mathrm{CE}(\mathfrak{h}_i)\to \mathrm{CE}(\mathfrak{h}_1\times_{\mathfrak{g}_2}\mathfrak{h}_2)$ are the inclusions of $\mathrm{CE}(\mathfrak{g})[\check{e}_{1},\tilde{e}_1]$ into $\mathrm{CE}(\mathfrak{g})[\check{e}_{1,1},\tilde{e}_1,\check{e}_{1,2}]$ given by $\check{e}_{1}\mapsto \check{e}_{1,i}$. Therefore, we have
\[
q_1^*b_2+q_2^*(-b_2)=(q_1^*-q_2^*)(\check{e}_1\tilde{e}_1)=(\check{e}_{1,1}-\check{e}_{1,2})\tilde{e}_1\;.
\]
As a consequence, the Fourier-Mukai transform $\Phi_{q_1^*b_2+q_2^*(-b_2)}$ acts on a degree $2k$ element $\omega=\sum_{n\in \mathbb{Z}}u^{k-n}(\alpha_{2n}+\check{e}_1\beta_{2n-1})$ in
$\mathrm{CE}(\mathfrak{g}_1)[[u^{-1},u]]$ as
\begin{align*}
\Phi_{q_1^*b_2+q_2^*(-b_2)}(\omega)&=\sum_{n\in \mathbb{Z}}u^{k-n}p_{2*}((1+u^{-1}(\check{e}_{1,1}-\check{e}_{1,2})\tilde{e}_1)(\alpha_{2n}+\check{e}_{1,1}\beta_{2n-1}))\\
&=\sum_{n\in \mathbb{Z}}u^{k-n}\pi_{1*}q_{2*}(\alpha_{2n}+\check{e}_{1,1}\beta_{2n-1}+u^{-1}(\check{e}_{1,1}-\check{e}_{1,2})\tilde{e}_1\alpha_{2n}-u^{-1}\check{e}_{1,2}\tilde{e}_1\check{e}_{1,1}\beta_{2n-1})\\
&=\sum_{n\in \mathbb{Z}}u^{k-n}\pi_{1*}(\beta_{2n-1}+u^{-1}\tilde{e}_1\alpha_{2n}-u^{-1}\check{e}_{1}\tilde{e}_1\beta_{2n-1})\\
&=\sum_{n\in \mathbb{Z}}u^{k-n-1}(\alpha_{2n}+\check{e}_{1}\beta_{2n-1})\\
&=u^{-1}\omega.
\end{align*}
The same holds for odd degree elements, so that $\Phi_{q_1^*b_2+q_2^*(-b_2)}=u^{-1}\mathrm{Id}$ and so $u\Phi_{-b_2}\circ \Phi_{b_2}=\mathrm{Id}$.
The same argument shows that  $\Phi_{b_2}\circ u\Phi_{-b_2}=\mathrm{Id}$, so that, finally, 
\[
\Phi_{b_2}^{-1}=u\Phi_{-b_2},
\]
i.e., we have shown that the Fourier-Mukai transform associated with a rational T-fold configuration is indeed invertible, with inverse provided (up to a shift in degree, given by the multiplication by $u$) by the Fourier-Mukai transform with opposite kernel 2-cochain. This completes the proof of the last statement in Section \ref{rational-tfold-config}.

\subsection{The case of SuperMinkowski space $\mathbb{R}^{1,8|\mathbf{16}+\mathbf{16}}$}
\label{Sec-Mink}
All of the above constructions  immediately generalize from $L_\infty$-algebras to super-$L_\infty$-algebras, and it is precisely in this more general setting that we find an interesting example from the string theory literature. 

\medskip
Let $\mathbf{16}$ be the unique irreducible real representation of $\mathrm{Spin}(8,1)$ and let $\{\gamma_a\}_{a = 0}^{d-1}$ be the corresponding Dirac representation on $\mathbb{C}^{16}$ of the Lorentzian $d= 9$ Clifford algebra. Write $\mathbf{16}+\mathbf{16}$ for the direct sum of two copies of the representation $\mathbf{16}$, and write
$
    \psi =\binom{\psi_1}{\psi_2}
$
  with $\psi_1$ and $\psi_2$  in $\mathbf{16}$ for an element $\psi$ in  $\mathbf{16}+\mathbf{16}$. Finally, for $a=0,\cdots,8$, consider the Dirac matrices
  \[
  \Gamma^{a}
  =
 \begin{pmatrix}
       0 & \gamma^a
       \\
       \gamma^a & 0
    \end{pmatrix} 
    \;, \qquad
   \Gamma_9^{\mathrm{IIA}}
  =
    \begin{pmatrix}
      0 & \mathrm{I}
      \\
      -\mathrm{I} & 0
    \end{pmatrix}
  \;,\qquad
   \Gamma_9^{\mathrm{IIB}}
     =
             \begin{pmatrix}
              0 & \mathrm{I}
              \\
              \mathrm{I} & 0
           \end{pmatrix}
          \,,
          \quad  \text{ and }\qquad 
           \Gamma_{10}
  =
    \begin{pmatrix}
      i \mathrm{I} & 0
      \\
      0 & -i \mathrm{I}
    \end{pmatrix}
  \,,
\]
where $I$ is the identity matrix. 
  The super-Minkowski super Lie algebra
  $
    \mathbb{R}^{8,1\vert \mathbf{16}+\mathbf{16}}
    $
  is the super Lie algebra whose dual Chevalley-Eilenberg algebra
  is the differential $(\mathbb{Z}, \mathbb{Z}/2)$-bigraded commutative algebra
  generated from elements $\{e^a\}_{a = 0}^{8}$ in bidegree $(1,\mathrm{even})$
  and from elements $\{\psi^\alpha\}_{\alpha = 1}^{32}$ in bidegree $(1,\mathrm{odd})$
  with differential given by
  $$
    d\psi^\alpha = 0
    \;\;\;\,,
    \;\;\;
    d e^a = \overline{\psi}  \Gamma^a \psi
    \,,
      $$
      where
  $
    \overline{\psi} \Gamma^a \psi
      =
    (C \Gamma^a)_{\alpha \beta} \, \psi^\alpha  \psi^\beta
  $,
with $C$ the charge conjugation matrix for the real representation $\mathbf{16}+\mathbf{16}$. Since $d\psi^\alpha=0$ for any $\alpha$, both 
\[
c_2^{\mathrm{IIA}}=\overline{\psi}\Gamma_9^{\mathrm{IIA}}\psi\qquad\text{ and }\qquad c_2^{\mathrm{IIB}}=\overline{\psi}\Gamma_9^{\mathrm{IIB}}
\]
are degree (2,even) cocycles on $\mathbb{R}^{8,1\vert \mathbf{16}+ \mathbf{16}}$. The central extensions they classify are obtained by adding a new degree (1,even) generator $e^9_A$ or $e^9_B$ to $\mathrm{CE}(\mathbb{R}^{8,1\vert \mathbf{16}+ \mathbf{16}})$ with differential
\[
de^9_A=\overline{\psi}\Gamma_9^{\mathrm{IIA}}\psi \qquad\text{ and }\qquad de^9_B=\overline{\psi}\Gamma_9^{\mathrm{IIB}}\psi \;,
\]
respectively. These two central extensions are, therefore, themselves super-Minkowski super Lie algebras. Namely, the extensions classified by $c_2^{\mathrm{IIA}}$ and $c_2^{\mathrm{IIB}}$ are
\[
\mathbb{R}^{9,1\vert \mathbf{16}+ \overline{\mathbf{16}}}\qquad\text{ and }\qquad \mathbb{R}^{9,1\vert \mathbf{16}+ \mathbf{16}},
\]
respectively.
Finally,  let $\mu_{F1}^{\mathrm{IIA}}$ be the degree (3,even) element in $\mathrm{CE}(\mathbb{R}^{9,1\vert \mathbf{16}+ \overline{\mathbf{16}}})$ given by
\[
\mu_{F1}^{\mathrm{IIA}}=\mu_{F1}^{8,1}-i\overline{\psi}  \Gamma_9^{\mathrm{IIA}} \Gamma_{10} \psi e^9_A= -i  \sum_{a=0}^8 \overline{\psi}  \Gamma_a \Gamma_{10} \psi   e^a- i\overline{\psi}  \Gamma_9^{\mathrm{IIA}} \Gamma_{10} \psi e^9_A\;.
\]
The element $\mu_{F1}^{\mathrm{IIA}}$ is actually a cocycle \cite{CAIB00}, so that
\[
d\mu_{F1}^{8,1}=(i\overline{\psi}  \Gamma_9^{\mathrm{IIA}} \Gamma_{10} \psi)(\overline{\psi}\Gamma_9^{\mathrm{IIA}}\psi)\;.
\]
A simple direct computation shows
$
 \Gamma_9^{\mathrm{IIB}} = i \, \Gamma_9^{\mathrm{IIA}} \Gamma_{10}
 $,
so that
\[
d\mu_{F1}^{8,1}= (\overline{\psi}  \Gamma_9^{\mathrm{IIB}}\psi)(\overline{\psi}\Gamma_9^{\mathrm{IIA}}\psi)=c_2^{\mathrm{IIA}}c_2^{\mathrm{IIB}}\;.
\]
As the element $\mu_{F1}^{8,1}$, as well as the elements $c_2^{\mathrm{IIA}}$ and $c_2^{\mathrm{IIB}}$ actually belong to the differential bigraded subalgebra $\mathrm{CE}(\mathbb{R}^{8,1\vert \mathbf{16}+ {\mathbf{16}}})$ of $\mathrm{CE}(\mathbb{R}^{9,1\vert \mathbf{16}+ \overline{\mathbf{16}}})$, the relation
\[
d\mu_{F1}^{8,1}=c_2^{\mathrm{IIA}}c_2^{\mathrm{IIB}}
\]
actually holds in $\mathrm{CE}(\mathbb{R}^{8,1\vert \mathbf{16}+ {\mathbf{16}}})$, so that the triple $(c_2^{\mathrm{IIA}},c_2^{\mathrm{IIB}},\mu_{F1}^{8,1})$ defines an $L_\infty$-morphism
\[
\mathbb{R}^{8,1\vert \mathbf{16}+ {\mathbf{16}}}
\longrightarrow
 b\mathfrak{tfold}.
\]
The 3-cocycles on $\mathbb{R}^{9,1\vert \mathbf{16}+ \overline{\mathbf{16}}}$ and on $\mathbb{R}^{9,1\vert \mathbf{16}+ {\mathbf{16}}}$ associated with this $L_\infty$-morphism are
\[
\mu_{F1}^{8,1}-e^9_Ac_2^{\mathrm{IIB}} \qquad \text{ and }\qquad\mu_{F1}^{8,1}-c_2^{\mathrm{IIA}}e^9_B,
\]
respectively. As $\Gamma_9^{\mathrm{IIB}} = i \, \Gamma_9^{\mathrm{IIA}} \Gamma_{10}$, we see that
\[
\mu_{F1}^{8,1}-e^9_Ac_2^{\mathrm{IIB}}=\mu_{F1}^{8,1}-e^9_A\overline\psi \Gamma_9^{\mathrm{IIB}}\psi=\mu_{F1}^{8,1}-i \overline\psi \Gamma_9^{\mathrm{IIA}} \Gamma_{10}\psi e^9_A=\mu_{F1}^{\mathrm{IIA}}\;.
\]
We then set $\mu_{F1}^{\mathrm{IIB}}=\mu_{F1}^{8,1}-c_2^{\mathrm{IIA}}e^9_B$. An explicit expression for the $(3,\mathrm{even})$-cocycle $\mu_{F1}^{\mathrm{IIB}}$ on $\mathbb{R}^{9,1\vert \mathbf{16}+ \overline{\mathbf{16}}}$ is
\[
\mu_{F1}^{\mathrm{IIB}}=\mu_{F1}^{8,1}-\overline\psi \Gamma_9^{\mathrm{IIA}}\psi e^9_B=-i  \sum_{a=0}^8\overline{\psi}  \Gamma_a \Gamma_{10} \psi   e^a
-i\overline{\psi}  \Gamma_9^{\mathrm{IIB}} \psi   e^9_B\;,
\]
where we used $\Gamma_9^{\mathrm{IIA}}=i\Gamma_9^{\mathrm{IIB}} \Gamma_{10}$. We have therefore an explicit Fourier-Mukai isomorphism
\[
\xymatrix{
\Phi_{e^9_Ae^9_B}\colon H_{L_\infty;\mu_{F1}^{\mathrm{IIA}}}^{\bullet}(\mathbb{R}^{9,1\vert \mathbf{16}+ \overline{\mathbf{16}}};\mathbb{R})[[u^{-1},u]]  
\; \ar[r]^-\sim & \;
H_{L_\infty;\mu_{F1}^{\mathrm{IIB}}}^{\bullet-1}(\mathbb{R}^{9,1\vert \mathbf{16}+ {\mathbf{16}}};\mathbb{R})[[u^{-1},u]]\;.
}
\]
This isomorphism is known as Hori's formula or as the Buscher rules for RR-fields in the string theory literature \cite{Ho}. A direct computation shows that it maps the 
$\mu_{F1}^{\mathrm{IIA}}$-twisted cocycles of \cite{CAIB00} on $\mathbb{R}^{9,1\vert \mathbf{16}+ \overline{\mathbf{16}}}$ to the 
$\mu_{F1}^{\mathrm{IIB}}$-twisted cocycles of \cite{IIBAlgebra} on $\mathbb{R}^{9,1\vert \mathbf{16}+ {\mathbf{16}}}$; see \cite{FSS16} for details.


\end{document}